\documentclass[twocolumn]{aastex701}

\usepackage{bm}
\usepackage{multirow}
\usepackage{amsmath}
\usepackage{color}
\usepackage{subfigure}
\usepackage{longtable}
\usepackage{booktabs}
\usepackage[T1]{fontenc}
\usepackage{lmodern}
\usepackage{tipx}

\newcommand{\hi}{\textsc{Hi}}
\newcommand{\hh}{\ensuremath{\rm H_2}}

\newcommand{\mhi}{\ensuremath{\textrm{M}_{\textsc{Hi}}}}
\newcommand{\msun}{\>{\rm M_{\odot}}}

\newcommand{\mstar}{\ensuremath{\textrm{M}_{\ast}}}

\newcommand{\lgfhi}{\ensuremath \log { M_{\hi}/M_*}}

\newcommand{\addd}{\ensuremath{\rm ATLAS^{3D}}}

\shorttitle{The \hi\ content of quenched massive galaxies}
\shortauthors{Li et al.}

\begin{document}

\title{What is the true H\,{\small I}\ gas content in massive quiescent galaxies in the local Universe?}

\correspondingauthor{Xiao Li}
\email{xli27938@gmail.com}
\correspondingauthor{Cheng Li}
\email{cli2015@tsinghua.edu.cn}

\author[0000-0002-2884-9781]{Xiao Li}\email{xli27938@gmail.com}
\affiliation{Department of Astronomy, Tsinghua University, Beijing 100084, China}
\affiliation{Max-Planck-Institut f\"ur Radioastronomie (MPIfR), Auf dem H\"ugel 69, 53121 Bonn, Germany}

\author[0000-0002-8711-8970]{Cheng Li}\email{cli2015@tsinghua.edu.cn}
\affiliation{Department of Astronomy, Tsinghua University, Beijing 100084, China}

\author[0000-0003-0202-0534]{Cheng  Cheng}\email{chengcheng@bao.ac.cn}
\affiliation{Chinese Academy of Sciences South America Center for Astronomy, National Astronomical Observatories, CAS, Beijing 100101, China}
\affiliation{CAS Key Laboratory of Optical Astronomy, National Astronomical Observatories, Chinese Academy of Sciences, Beijing 100101, China}

\author[0000-0001-5356-2419]{H. J. Mo}\email{hjmo@umass.edu}
\affiliation{Department of Astronomy, University of Massachusetts Amherst, MA 01003, USA}

\author[0000-0002-6593-8820]{Jing Wang}\email{jwang_astro@pku.edu.cn}
\affiliation{Kavli Institude for Astronomy and Astrophysics, Peking University, Beijing 100871, China}

\author[0000-0003-4357-3450]{Am\'elie Saintonge}
\email{asaintonge@mpifr-bonn.mpg.de}
\affiliation{Max-Planck-Institut f\"ur Radioastronomie (MPIfR), Auf dem H\"ugel 69, 53121 Bonn, Germany}
\affiliation{Department of Physics and Astronomy, University College London, Gower Street, London WC1E 6BT, UK}

\begin{abstract}
While massive quiescent galaxies are known to be poor in atomic hydrogen (\hi), their true \hi\ content remains poorly constrained due to the limited sensitivity and morphological biases of existing surveys. We present deep \hi\ observations using the Five-hundred-meter Aperture Spherical radio Telescope (FAST) for a representative sample of 78 low-redshift massive quiescent galaxies, selected by stellar mass ($M_\ast > 10^{10} M_\odot$), color (NUV$-r > 5$), and specific star formation rate ($\rm \log sSFR < -11\ yr^{-1}$). Our observations reach a remarkable sensitivity of $\log(M_{\textsc{Hi}}/M_\ast) = -2.6$ for 55 targets and $\log(M_{\textsc{Hi}}/M_\ast) = -3.2$ for 23 targets. We find that one-third of the sample follows the \hi\ scaling relation derived from previous surveys, while the remaining two-thirds exhibit significantly lower \hi\ content. 
The \hi\ mass fraction shows no clear correlation with specific star formation rate, NUV$-r$ color index, stellar surface mass density, and concentration. Our FAST sample shows remarkable similarity to the \addd\ sample which only includes early-type galaxies, both in its high fraction of \hi-poor galaxies and its high satellite fraction among \hi-poor galaxies. 
These results suggest that while both early-type morphology and environment may contribute to the extreme \hi\ deficiency, neither factor alone fully explains the observed gas depletion, indicating that additional physical mechanisms must be responsible for the extreme \hi\ deficiency prevalent in massive quiescent galaxies.
\end{abstract}

\section{Introduction}
\label{sec:intro}

\begin{figure*}
    \centering
    \includegraphics[width=0.8\textwidth]{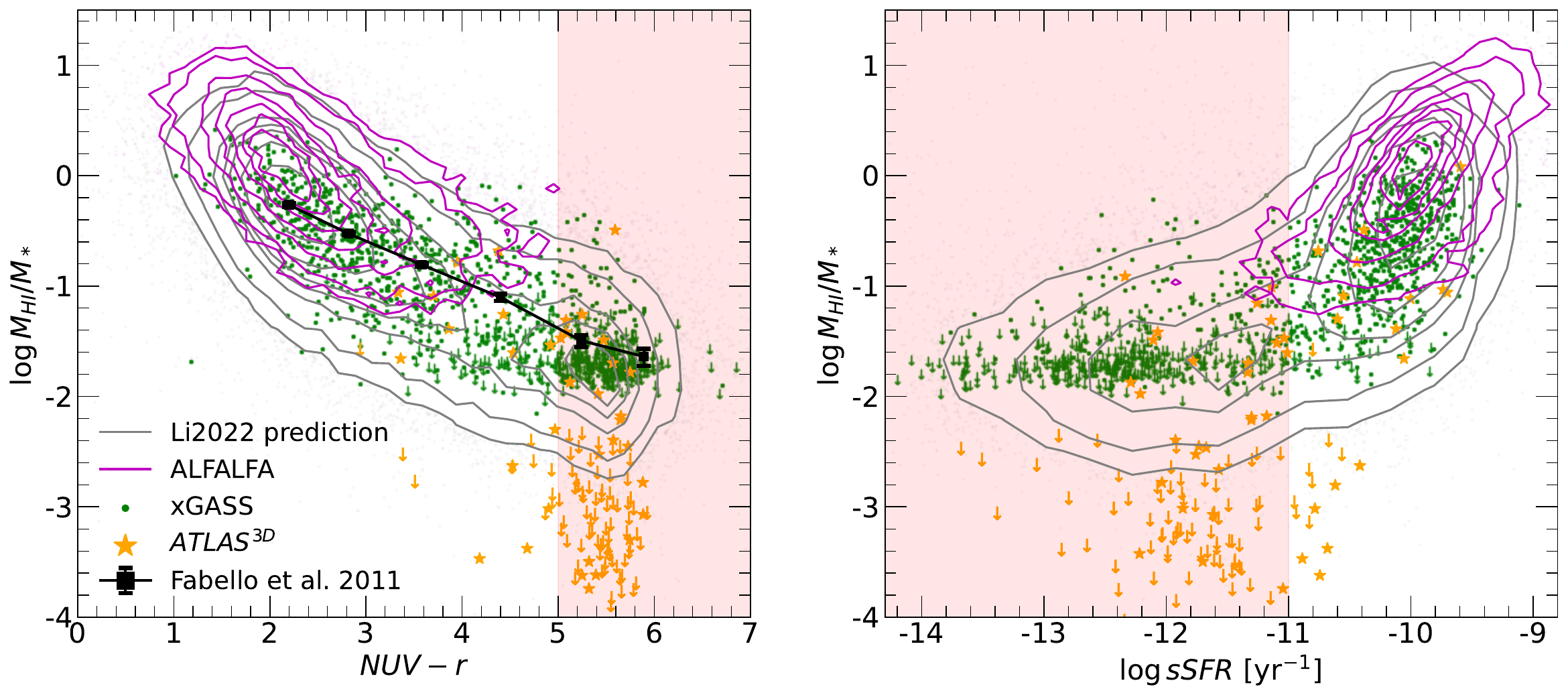}
    \caption{\hi\ mass fraction as a function of NUV$-r$ (left) and specific star formation rate (right). The gray contours represent the SDSS volume-limited sample. The magenta contours represent the ALFALFA 100 percent sample. The green points indicate the xGASS representative sample. The orange stars denote the \addd\ sample. \hi\ upper limits are shown as downward arrows. The red regions highlight the parameter space of quiescent galaxies.}
    \label{fig:intro}
\end{figure*}

The galaxy population is well known to exhibit a bimodal distribution, consisting of a star-forming main sequence and a quiescent population \citep[e.g.][]{Baldry2004,Kauffmann2004,Schawinski2014}. Extensive atomic hydrogen (\hi) surveys have shown that star-forming galaxies possess substantial \hi\ reservoirs, with gas fractions ($M_{\hi}/M_*$) typically ranging from 0.1 to 10 \citep{Kannappan2004,GASS,HuangS2012,xGASS}. In contrast, the majority of quiescent galaxies remain undetected in these surveys, exhibiting significantly lower \hi\ fractions of $M_{\hi}/M_\ast \lesssim 1\%$ \citep{Serra2012-ATLAS3D,xGASS}. This paucity of cold gas extends to the molecular phase, as most quiescent galaxies also lack detectable CO emission, with molecular hydrogen fractions $M_{\hh}/M_\ast \lesssim 1\%$ \citep{Saintonge2011a,Saintonge2017ApJS-xCOLDGASS,Young2011,Bothwell2014-ALLSMOG}. Given that star formation is fueled by cold gas (\hi\ and \hh), these observations support the consensus that galaxies stop their star formation activity mostly due to the depletion or absence of their gas reservoirs \citep{Cortese2020,Guo2021,Saintonge2022}. Although a rare population of quiescent yet \hi-rich galaxies does exist \citep[e.g.][]{Lemonias2014,XiaoLi2024,LiFujia2024,LiYang2025}, recent work confirms that they are exceptionally uncommon \citep{XiaoLi2024}.

Although it is well established that quiescent galaxies do not have large \hi\ gas reservoirs, their precise \hi\ content and its dependence on galaxy properties remain poorly constrained, limited by the sensitivity of existing \hi\ surveys. This limitation is evident in \autoref{fig:intro}: the \hi-detected galaxies from the ALFALFA survey \citep[magenta contours;][]{Giovanelli_2005,ALFALFA100} are dominated by blue, star-forming systems with near-ultraviolet (NUV) to optical color NUV$-r\lesssim 5$ and specific star formation rate $\text{sSFR} \gtrsim 10^{-11}~\text{yr}^{-1}$, reaching a detection limit of  $M_{\textsc{Hi}}/M_\ast\sim5\%$. The xGASS survey \citep[][]{GASS,xGASS}, though with a much smaller sample, has achieved a significantly lower detection limit of $M_{\textsc{Hi}}/M_\ast\sim1.5\%$ for a representative sample spanning the full range of NUV$-r$ color and sSFR (see the green points/arrows in~\autoref{fig:intro}). Pushing these limits further, the \addd\ sample \citep[][]{Cappellari2011-ATLAS3D,Serra2012-ATLAS3D} obtained deep \hi\ observations for 166 early-type galaxies (orange stars/arrows in~\autoref{fig:intro}), reaching mass fraction limits as low as $M_{\textsc{Hi}}/M_*\sim10^{-4}$. This sample, though primarily yielding upper limits, definitively shows that early-type galaxies are overwhelmingly \hi-poor. Its focus on this specific morphological type, however, means it cannot fully represent the general population of massive quiescent galaxies.

To overcome the depth limitations of existing \hi\ surveys, stacking analyses have been employed. This technique co-adds \hi\ observations of individual galaxies with similar properties, achieving the sensitivity required to detect their average \hi\ content \citep[e.g.,][]{Fabello2011,Fabello2011b,Fabello2012,Gereb2013,Brown2015,Guo-HI-halo-relation,Namiki2021,Bianchetti2025}. For instance, \citet{Fabello2011} stacked ALFALFA spectra to estimate average \hi\ mass fractions as a function of galaxy color and structural parameters. They found that $M_{\textsc{Hi}}/M_\ast$ correlates most strongly with NUV$-r$ color and surface stellar mass density ($\mu_\ast$), but shows no dependence on the presence of a prominent bulge when their color or surface density is controlled. \autoref{fig:intro} shows this $M_{\textsc{Hi}}/M_\ast$ versus NUV$-r$ relation from their work. In a more recent study, \citet{Namiki2021} also used ALFALFA stacking to investigate the relationship between \hi\ content, stellar mass, and star formation rate in star-forming galaxies. While they similarly found no clear dependence on broad morphological type, for early-type galaxies they suggested that the presence of small-scale structures may be linked to their total \hi\ content. Typically, stacking detections of massive quiescent galaxies reach mass fractions of $M_{\textsc{Hi}}/M_\ast\sim2-3\%$, far above the very low \hi\ fraction prevalent in early-type massive quiescent galaxies as revealed by the deeper \addd\ survey, and the information of the \hi\ fraction distribution is lost.

As an alternative approach, scaling relations have been used to estimate the \hi\ gas mass fractions of optically selected galaxies based on their spectroscopic and photometric properties \citep[e.g.,][]{Kannappan2004,Tremonti2004,Kannappan2008,ZhangWei_2009,GASS,Li2012,2015ApJ...810..166E,Zu2020,XiaoLi,Lu2024}. In \citet{XiaoLi}, we developed a estimator that relates $\log (M_{\textsc{Hi}}/M_\ast)$ to a linear combination of four parameters: the $u-r$ color index, stellar surface density $\mu_\ast$, stellar mass $M_\ast$, and the concentration index. The scatter of individual galaxies around the mean relation is modeled by a Gaussian distribution. 
Calibrated with both the xGASS sample and the \hi\ mass function measured by ALFALFA using Bayesian inference, this estimator can reproduce the observed \hi\ mass function, \hi\ scaling relations, and the observed \hi-detected galaxy population for the xGASS and ALFALFA sample. We assume this estimator as a universal relation and use it to predict the \hi\ content of a volume-limited Sloan Digital Sky Survey \citep[SDSS;][]{York2000} sample with $M_\ast > 10^{9} M_{\odot}$ and $0.01 < z < 0.035$, shown as the gray contours in ~\autoref{fig:intro}. 
For quiescent galaxies (NUV$-r > 5$ or $\text{sSFR} < 10^{-11}~\text{yr}^{-1}$), the predicted \hi\ mass fraction reaches down to $M_{\textsc{Hi}}/M_\ast \sim 0.3\%$, nearly an order of magnitude lower than the detection limits of xGASS and stacking analyses. However, this predicted distribution does not encompass most \addd\ galaxies, indicating that the estimator fails to capture the \hi\ properties of those extremely \hi-poor galaxies, likely due to the limited depth of the calibration sample. Although the \addd\ sample has extraordinary $M_{\hi}/M_*$ detection limit, it is not representative of the massive quiescent galaxy population due to its morphology-selected nature.

Therefore, to determine the true \hi\ content of massive quiescent galaxies, it is essential to obtain deep \hi\ observations for a representative sample selected without morphological bias. Motivated by this need, we conducted deep \hi\ observations with the Five-hundred-meter Aperture Spherical radio Telescope \citep[FAST;][]{Nan2006-FAST,Jiang2019-FAST,Jiang2020-FAST} for a sample of 78 massive quiescent galaxies ($M_* > 10^{10} M_\odot$, NUV$-r > 5$). Our observations reach sensitivities of $\log(M_{\textsc{Hi}}/M_*) = -2.6$ for all galaxies and $\log(M_{\textsc{Hi}}/M_*) = -3.2$ for a subset. Selected solely by stellar mass and color, this sample is representative of the general massive quiescent population. In this paper, we present the results from our FAST observations. We incorporate the \addd\ and xGASS samples for a comprehensive comparison: the former is deep but morphologically biased, while the latter is shallower but representative; both provide valuable constraints. Our study aims to address the following questions: (1) What is the true \hi\ content of massive quiescent galaxies in the local universe? (2) Is the unusually low  \hi\ content as seen for the majority of early-type galaxies in \addd\ a general property of all massive quiescent galaxies in the local Universe? (3) If yes, what causes the extreme \hi\ poverty of these galaxies, and particularly, what are the roles of morphology and environment?

The paper is structured as follows. Section \ref{sec:data} describes the selection of our FAST, xGASS, and \addd\ samples. Section \ref{sec:results} presents our results. We discuss our findings in section \ref{sec:discussion} and summarize in section \ref{sec:summary}. Details of the FAST data reduction are provided in \autoref{app:data_reduction}. A $\Lambda$CDM cosmology with $(\Omega_m, \Omega_\Lambda, h) = (0.3, 0.7, 0.7)$ is adopted throughout this work.

\section{Data}\label{sec:data}


\begin{figure*}[ht]
    \centering
    \includegraphics[height=0.4\linewidth]{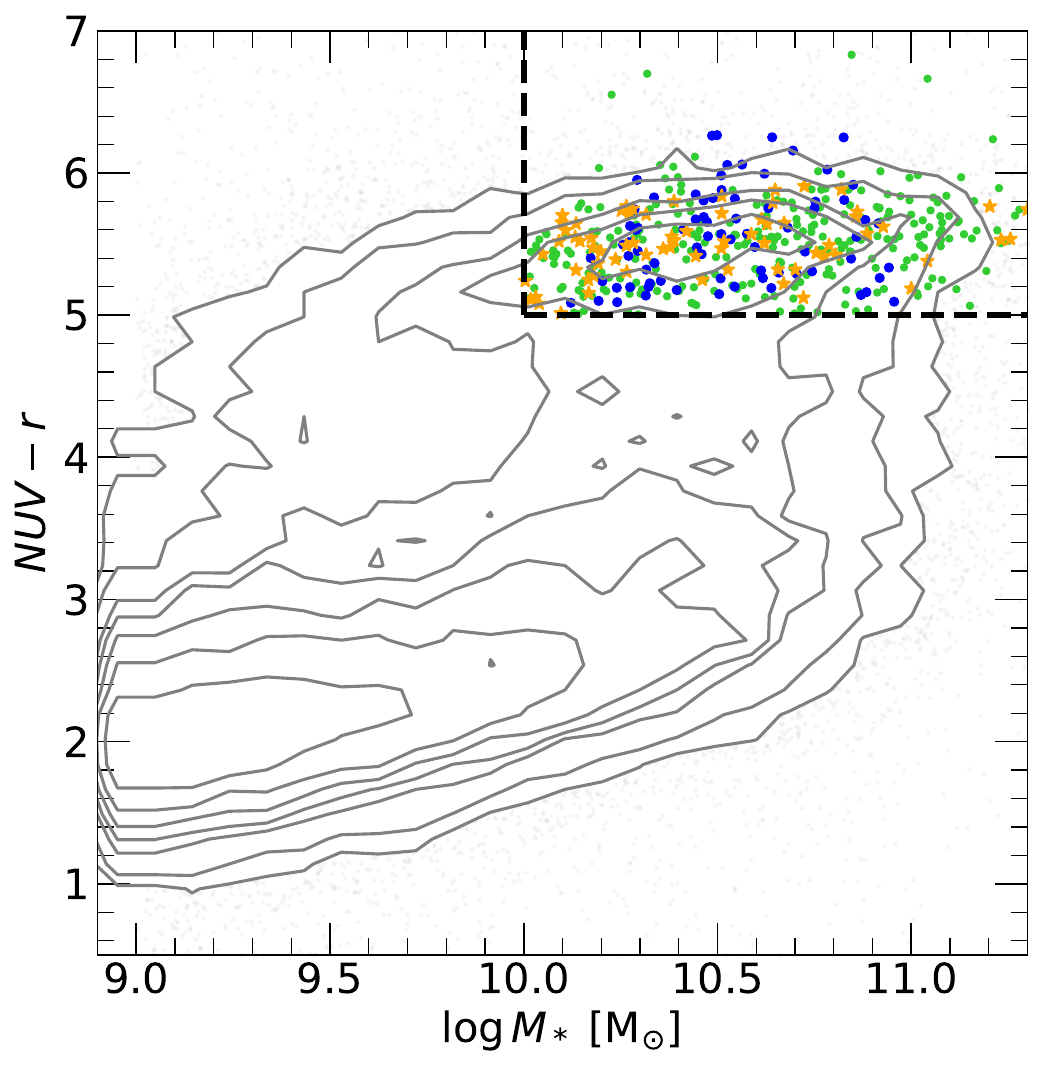}
    \hspace{0.3cm}
    \includegraphics[height=0.4\linewidth]{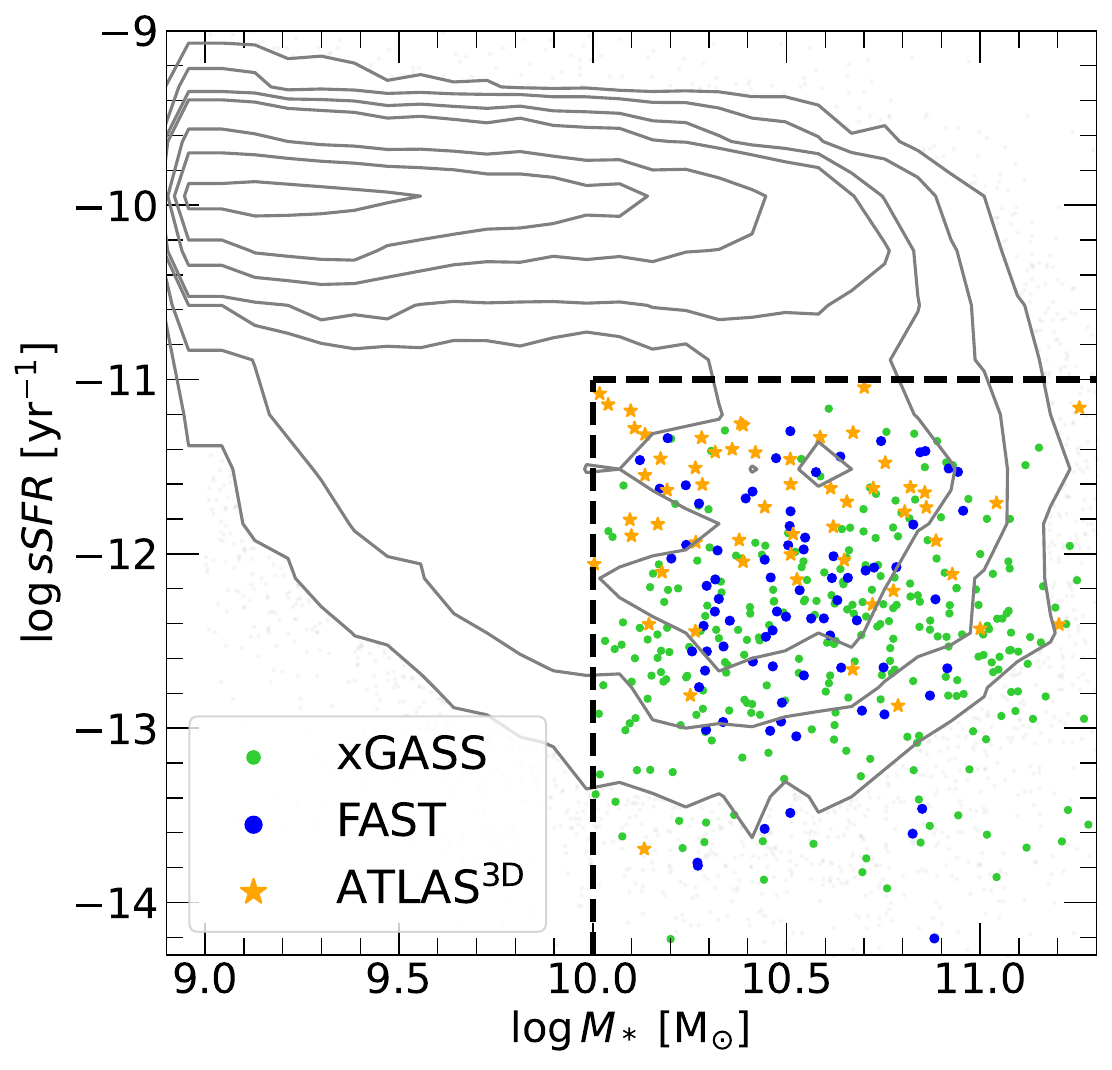}    
    \caption{NUV$-r$ (left) and Specific star formation rate (right) versus stellar mass. The gray contours represent the SDSS volume-limited sample. Contours from innermost to outermost include 15\%, 30\%, 45\%, 60\%, 75\%, 90\%, and 95\% of the volume-limited sample, respectively. The colored symbols indicate the \hi\ samples (xGASS: green dots; FAST: blue dots; \addd: orange stars). The black dashed lines represent our sample selection criteria (vertical: $M_*>10^{10} M_{\odot}$, horizontal: $\log {\rm sSFR} < -11$ or NUV$-r > 5$).}
    \label{fig:SFR_Mstar}
\end{figure*}

\subsection{The FAST sample and \hi\ observations} \label{subsec:sample}

Our FAST sample is selected from the NASA-Sloan Atlas (NSA; \citealt{Blanton_2011})\footnote{\url{http://nsatlas.org}}, which includes $\sim 640,000$ galaxies out to $z < 0.15$ from SDSS DR8 \citep{SDSS-DR8}.
Our selection criteria include: $+10^{\circ} < \text{Dec} < +40^{\circ}$, $0.01 < z < 0.02$, $M_* > 10^{10} M_\odot$, $\text{NUV}-r > 5$, and $\rm \log sSFR < -11\ yr^{-1}$. Stellar masses and $\text{NUV}-r$ colors are taken directly from the NSA catalog. The stellar masses were derived by fitting the spectral energy distribution (SED) to GALEX \citep{Martin_2005} FUV, NUV, and SDSS $ugriz$ photometry, assuming a Chabrier initial mass function \citep{Chabrier2003}. The $\text{NUV}$- and $r$-band magnitudes were measured using an elliptical Petrosian aperture model and are corrected for Galactic extinction. The sSFR is estimated by performing spectral energy distribution (SED) fitting from the ultraviolet to the infrared using the \texttt{CIGALE} code \citep{cigale2019, cigale2020}. The UV-to-optical photometry was taken from the NSA, which provides elliptical Petrosian magnitudes in FUV, NUV, $u, g, r, i$ and $z$ band. Infrared photometry was incorporated using W3 and W4 magnitudes from the unWISE catalog \citep{unWISE_Lang}, which performs forced photometry at SDSS galaxy positions on WISE coadds.
The declination constraint ensures that the target galaxies are observed at small zenith angles (ZA $< 25^{\circ}$), where FAST's performance is optimal. The chosen redshift range places the sample in the nearby universe, which is necessary to achieve our intended detection limits. The cuts in stellar mass and $\text{NUV}-r$ color ensure we target massive quiescent galaxies. Applying these criteria resulted in a parent sample of 413 galaxies. From this parent sample, we randomly selected 78 galaxies for our FAST observations, based on the available telescope time. According to the morphology classification scheme described in \cite{Masters2025}, this sample comprises 7(9\%) late-type galaxies and 71(91\%) early-type galaxies. As we will demonstrate in \autoref{result:morphology}, this sample is representative of the massive quiescent population in terms of morphology. 
We checked the radio continuum sky map from the VLASS survey\footnote{https://www.legacysurvey.org} \citep{VLASS}, and found all except 4 of the target galaxies do not have any radio continuum source within the FAST beam at either the on-source or off-source positions. We did not remove any target galaxy in this step. The effects of radio continuum sources were accounted for in the error estimation (see \autoref{app:data_reduction} for details).
We confirmed that all target galaxies are free of \hi\ confusion by ensuring no companion galaxies exist within a projected angular separation of $\theta < 3^\prime$ (the beam size of FAST) and a radial velocity offset of $|v| < 500~\text{km~s}^{-1}$ (derived from optical redshifts). 
We also inspected the Legacy Survey images to see if there are any faint neighbor galaxies that are missed by the SDSS spectroscopic survey. Among our target galaxies, we identified eight that possess at least one neighboring galaxy within an angular separation of $3'$. As these neighboring galaxies lack redshift measurements, we cannot ascertain whether they contribute to \hi\ confusion. Consequently, we retained all eight galaxies in the FAST target sample. We have verified that excluding them does not alter our conclusions.

Figure~\ref{fig:SFR_Mstar} displays the distribution of our FAST sample (blue dots) on the NUV$-r$ versus $M_\ast$ and sSFR versus $M_\ast$ diagrams (left and right panels, respectively). The black dashed lines in the left panel indicate our sample selection criteria. For reference, the predicted distribution of the SDSS volume-limited sample from the previous figure is shown as gray contours in both panels. For this volume-limited sample, stellar masses are also taken from the NSA, and the sSFR is estimated using the same method as for the FAST sample.  
As shown in the right panel, all galaxies in our FAST sample fall in the ``massive quiescent region'' defined by $\log(M_\ast/M_\odot) > 10$, $\log\text{sSFR} < -11\ {\rm yr^{-1}}$, and NUV$-r > 5$ (dashed lines).

The first round of \hi\ observations was conducted from September 2023 to June 2024. We observed 74 galaxies to a sensitivity of $\log(M_{\textsc{Hi}}/M_\ast) = -2.6$ and 4 galaxies to $\log(M_{\textsc{Hi}}/M_\ast) = -3.2$ (both at $3\sigma$), assuming a characteristic \hi\ line width of 300 km s$^{-1}$, which is a good estimation of the rotation velocity or velocity dispersion of massive ($M_* > 10^{10}\msun$) galaxies \citep{Veale2018,Lelli2019}. 
Since a galaxy is known to exist at the target position, we employed a $3\sigma$ criterion for detection.
This initial campaign yielded 31 detections and 47 non-detections. The high fraction of non-detections is not unexpected, given both the similarly high fraction of non-detections in the \addd\ sample and the fact that the majority of massive quiescent galaxies are early-type. To further constrain the true \hi\ content of our sample, we performed a second round of deeper observations targeting 19 galaxies from September to October 2024, reaching a sensitivity of $\log(M_{\textsc{Hi}}/M_\ast) = -3.2$ ($3\sigma$). The 19 targets were selected from the initial non-detections that required the least observation time, so that we can get a deep sample as large as possible within the limited observation time. Since the whole FAST sample has a narrow range of stellar mass and redshift, this selection method is supposed to introduce negligible bias to the final \hi\ sample. This follow-up resulted in 3 new detections and 16 non-detections. Consequently, our final FAST sample consists of 34 \hi\ detections and 44 non-detections. The basic properties and \hi\ observational results for the sample are listed in \autoref{table:galaxylist}. Details of the observational setup and data reduction are provided in~\autoref{app:data_reduction}.

Among the 78 targets in our FAST sample, six galaxies had prior \hi\ detections from ALFALFA. We retained these in our sample to enable a consistency check between the telescopes. A detailed comparison of the \hi\ fluxes, masses, and spectra derived from FAST and ALFALFA observations is presented in~\autoref{app:data_reduction}. Our analysis reveals a systematic offset, with FAST measurements yielding lower integrated fluxes than those from ALFALFA. While the exact cause of this discrepancy remains uncertain, evidence suggests that it may stem from differences in observational methodology: our single-pointing FAST observations are limited to the \hi\ gas within the beam area, while ALFALFA's mapping mode can capture the extended \hi\ gas in the galaxy outskirts.
Interestingly, for low signal-to-noise ratio ($\rm SNR\lesssim20$) sources, \cite{FASHI} also reported systematic lower \hi\ fluxes  from FAST compared to ALFALFA, suggesting that there may be an intrinsic bias between FAST and ALFALFA measurements. We note that while this systematic effect may introduce quantitative differences in absolute flux measurements, it does not affect the qualitative conclusions of this study regarding relative \hi\ content and population trends.

\renewcommand{\arraystretch}{0.89}
\begin{longtable}{ccccccccccccc}
    \caption{Basic information of the FAST sample.\label{table:galaxylist}}\\
    \hline
    \multicolumn{1}{c}{(1)} & \multicolumn{1}{c}{(2)} & \multicolumn{1}{c}{(3)} & \multicolumn{1}{c}{(4)} & \multicolumn{1}{c}{(5)} & \multicolumn{1}{c}{(6)} & \multicolumn{1}{c}{(7)} & \multicolumn{1}{c}{(8)} & \multicolumn{1}{c}{(9)} & \multicolumn{1}{c}{(10)} & \multicolumn{1}{c}{(11)} & \multicolumn{1}{c}{(12)}\\
    \multicolumn{1}{c}{ID} & \multicolumn{1}{c}{Source Name} & \multicolumn{1}{c}{R.A.} & \multicolumn{1}{c}{decl.} & \multicolumn{1}{c}{$z$} & \multicolumn{1}{c}{$\log M_*$} & \multicolumn{1}{c}{NUV$-r$} & \multicolumn{1}{c}{$\rm \log sSFR$} & \multicolumn{1}{c}{T-type} & \multicolumn{1}{c}{$\log M_{\hi}$} & S/N & \multicolumn{1}{c}{$W_{\rm \hi }$} \\
    \multicolumn{1}{c}{} & \multicolumn{1}{c}{} & \multicolumn{1}{c}{[deg]} & \multicolumn{1}{c}{[deg]} & \multicolumn{1}{c}{} & \multicolumn{1}{c}{[$\rm M_{\odot}$]} & \multicolumn{1}{c}{[mag]} & \multicolumn{1}{c}{[$\rm yr^{-1}$]} & \multicolumn{1}{c}{} & \multicolumn{1}{c}{[$\rm M_{\odot}$]} & \multicolumn{1}{c}{} & \multicolumn{1}{c}{[$\rm km/s$]} \\
    \hline
    \endfirsthead %
    \multicolumn{12}{c}{{\tablename\ \thetable{} -- continued}} \\
    \hline
    \multicolumn{1}{c}{(1)} & \multicolumn{1}{c}{(2)} & \multicolumn{1}{c}{(3)} & \multicolumn{1}{c}{(4)} & \multicolumn{1}{c}{(5)} & \multicolumn{1}{c}{(6)} & \multicolumn{1}{c}{(7)} & \multicolumn{1}{c}{(8)} & \multicolumn{1}{c}{(9)} & \multicolumn{1}{c}{(10)} &
    \multicolumn{1}{c}{(11)} & \multicolumn{1}{c}{(12)}\\
    ID & Source Name & R.A. & decl. & $z$ & $\log M_*$ & NUV$-r$ & $\rm \log sSFR$ & T-type & $\log M_{\hi}$ & S/N & $W_{\rm \hi }$ \\
    \multicolumn{1}{c}{} & \multicolumn{1}{c}{} & \multicolumn{1}{c}{[deg]} & \multicolumn{1}{c}{[deg]} & \multicolumn{1}{c}{} & \multicolumn{1}{c}{[$\rm M_{\odot}$]} & \multicolumn{1}{c}{[mag]} & \multicolumn{1}{c}{[$\rm yr^{-1}$]} & \multicolumn{1}{c}{} & \multicolumn{1}{c}{[$\rm M_{\odot}$]} & \multicolumn{1}{c}{} & \multicolumn{1}{c}{[$\rm km/s$]} \\
    \hline
    \endhead %
    \hline
    \endfoot %
    \hline
    \vspace{0.1cm}
    \\
    \multicolumn{12}{l}{\footnotesize \parbox{0.95\linewidth}{%
            Note: (1) Source ID, (2) Source name, (3) R.A., (4) decl., (5) optical redshift, (6) stellar mass, (7) NUV$-r$ color index, (8) specific star formation rate, (9) morphology T-type, (10) \hi\ mass, (11) signal-to-noise ratio of \hi\ mass, (12) \hi\ velocity width defined by the width of velocity range over which the integrated \hi\ flux is measured. The uncertainty of velocity width is estimated by the Monte Carlo method following \citep{YuNiankun2020}. For sources without \hi\ detections, we report $3\sigma$ upper limits on their \hi\ masses. The stellar mass and NUV$-r$ color index are taken from the NASA-Sloan Atlas \citep{Blanton_2011}. Velocities and redshifts are heliocentric.\\
            $\dagger:$ The reported \hi\ mass of UGC 1939 (ID=24) should be treated with caution. See \autoref{app:data_reduction} for details.}
        } \\
    \endlastfoot 

    1 & UGC 130         & 3.4871 & 30.8829 & 0.0160 & 10.46 & 5.43 & -12.09 & -5 & 8.20$\pm $ 0.11 & 4.6 & 328$\pm $104 \\
2 & LEDA 1287       & 5.0082 & 21.9998 & 0.0198 & 10.50 & 5.15 & -12.07 & -5 & $<$ 7.96 & - & - \\
3 & NGC 243         & 11.5036 & 29.9596 & 0.0160 & 10.51 & 5.26 & -12.08 & -5 & 8.56$\pm $ 0.05 & 23.2 & 270$\pm $26 \\
4 & UGC 636         & 15.4256 & 24.0579 & 0.0170 & 10.62 & 5.26 & -12.22 & -5 & $<$ 7.56 & - & - \\
5 & UGC 646         & 15.8599 & 32.2374 & 0.0176 & 10.58 & 5.57 & -11.57 & -1 & 8.87$\pm $ 0.07 & 16.2 & 367$\pm $18 \\
6 & LEDA 3881       & 16.3925 & 32.4297 & 0.0168 & 10.34 & 5.37 & -12.60 & -5 & $<$ 8.00 & - & - \\
7 & NGC 388         & 16.9465 & 32.3099 & 0.0182 & 10.27 & 5.51 & -14.14 & -5 & $<$ 7.45 & - & - \\
8 & NGC 398         & 17.2237 & 32.5145 & 0.0166 & 10.33 & 5.22 & -12.27 & -5 & $<$ 7.82 & - & - \\
9 & NGC 399         & 17.2467 & 32.6342 & 0.0172 & 10.75 & 5.75 & -12.70 & -1 & $<$ 8.23 & - & - \\
10 & LEDA 4244       & 17.7972 & 32.7018 & 0.0169 & 10.51 & 5.98 & -13.53 & -2 & $<$ 7.32 & - & - \\
11 & NGC 431         & 18.5189 & 33.7043 & 0.0191 & 10.94 & 5.33 & -11.61 & -2 & $<$ 7.79 & - & - \\
12 & NGC 447         & 18.9069 & 33.0678 & 0.0187 & 10.84 & 5.40 & -11.47 & 5 & 9.50$\pm $ 0.05 & 55.1 & 212$\pm $7 \\
13 & NGC 472         & 20.1196 & 32.7091 & 0.0177 & 10.73 & 5.45 & -12.21 & -5 & 8.37$\pm $ 0.09 & 6.0 & 328$\pm $67 \\
14 & LEDA 4910       & 20.3228 & 33.0904 & 0.0171 & 10.32 & 5.20 & -12.03 & -5 & $<$ 7.70 & - & - \\
15 & UGC 901         & 20.4055 & 32.6060 & 0.0158 & 10.27 & 5.63 & -12.86 & -2 & $<$ 7.58 & - & - \\
16 & NGC 501         & 20.8434 & 33.4329 & 0.0167 & 10.29 & 5.45 & -12.40 & -5 & $<$ 7.64 & - & - \\
17 & NGC 503         & 20.8684 & 33.3318 & 0.0198 & 10.49 & 5.82 & -13.18 & -3 & 7.94$\pm $ 0.11 & 4.9 & 328$\pm $96 \\
18 & LEDA 5129       & 20.9938 & 33.3132 & 0.0168 & 10.29 & 5.95 & -13.19 & -1 & $<$ 7.70 & - & - \\
19 & LEDA 5951       & 24.0845 & 17.8162 & 0.0197 & 10.32 & 5.32 & -12.34 & -5 & 8.43$\pm $ 0.07 & 14.9 & 251$\pm $43 \\
20 & UGC 1271        & 27.2503 & 13.2111 & 0.0168 & 10.75 & 5.80 & -13.03 & 3 & $<$ 7.65 & - & - \\
21 & NGC 711         & 28.1158 & 17.5127 & 0.0164 & 10.64 & 5.19 & -11.57 & -2 & 9.20$\pm $ 0.05 & 34.7 & 444$\pm $23 \\
22 & UGC 1496        & 30.2491 & 15.1892 & 0.0155 & 10.66 & 5.30 & -12.03 & -1 & 8.88$\pm $ 0.07 & 12.0 & 309$\pm $76 \\
23 & UGC 1512        & 30.4631 & 16.2578 & 0.0152 & 10.53 & 6.06 & -13.17 & -3 & $<$ 7.55 & - & - \\
24 & UGC 1939        & 37.0733 & 26.3125 & 0.0174 & 10.63 & 5.75 & -12.40 & 10 & 10.00$\pm $ 0.05 & 204.4 & 560$\pm $2 \\
25 & LEDA 9925       & 39.2453 & 25.4429 & 0.0184 & 10.85 & 5.92 & -13.54 & -5 & $<$ 7.76 & - & - \\
26 & UGC 2218        & 41.2408 & 32.7063 & 0.0177 & 10.70 & 5.29 & -12.16 & -5 & 8.39$\pm $ 0.07 & 8.4 & 386$\pm $38 \\
27 & LEDA 3088276    & 54.5706 & 37.8549 & 0.0187 & 10.92 & 5.26 & -11.56 & -2 & $<$ 7.82 & - & - \\
28 & LEDA 19954      & 104.6977 & 29.0096 & 0.0148 & 10.49 & 6.26 & -12.96 & -5 & 8.52$\pm $ 0.1 & 5.2 & 405$\pm $83 \\
29 & NGC 2379        & 111.8594 & 33.8114 & 0.0134 & 10.44 & 5.67 & -13.72 & -5 & 7.56$\pm $ 0.1 & 4.7 & 251$\pm $112 \\
30 & NGC 2411        & 113.6514 & 18.2816 & 0.0169 & 10.83 & 6.25 & -13.67 & -5 & $<$ 8.33 & - & - \\
31 & LEDA 21372      & 114.0700 & 33.1227 & 0.0157 & 10.51 & 5.88 & -11.75 & -1 & 8.65$\pm $ 0.08 & 9.1 & 425$\pm $68 \\
32 & LEDA 22460      & 120.1530 & 15.6842 & 0.0157 & 10.41 & 5.60 & -11.63 & -5 & $<$ 7.91 & - & - \\
33 & UGC 4170        & 120.3464 & 15.3693 & 0.0156 & 10.78 & 6.02 & -12.10 & -5 & 9.02$\pm $ 0.05 & 33.4 & 637$\pm $31 \\
34 & LEDA 23044      & 123.3264 & 34.9325 & 0.0174 & 10.35 & 5.24 & -12.43 & -1 & $<$ 7.98 & - & - \\
35 & LEDA 23204      & 124.1412 & 21.4099 & 0.0156 & 10.61 & 5.31 & -12.49 & -5 & $<$ 7.61 & - & - \\
36 & NGC 2553        & 124.3959 & 20.9031 & 0.0156 & 10.69 & 6.16 & -12.92 & -1 & $<$ 7.68 & - & - \\
37 & LEDA 25080      & 133.9687 & 18.1565 & 0.0141 & 10.17 & 5.40 & -11.60 & -5 & 7.58$\pm $ 0.11 & 5.1 & 38$\pm $13 \\
38 & NGC 2737        & 135.9988 & 21.9066 & 0.0105 & 10.27 & 5.42 & -13.71 & -2 & 8.65$\pm $ 0.07 & 15.7 & 289$\pm $64 \\
39 & LEDA 27225      & 143.6698 & 19.5503 & 0.0147 & 10.29 & 5.61 & -12.53 & -3 & 8.66$\pm $ 0.07 & 10.2 & 502$\pm $96 \\
40 & LEDA 30617      & 156.4488 & 26.5707 & 0.0168 & 10.60 & 5.48 & -12.46 & -3 & $<$ 7.78 & - & - \\
41 & LEDA 30979      & 157.5873 & 15.1922 & 0.0191 & 10.54 & 5.20 & -12.24 & -5 & $<$ 7.83 & - & - \\
42 & LEDA 1630609    & 175.8538 & 20.5576 & 0.0189 & 10.12 & 5.09 & -11.53 & -5 & $<$ 7.63 & - & - \\
43 & LEDA 36465      & 175.9982 & 19.7789 & 0.0187 & 10.24 & 5.19 & -11.97 & -2 & $<$ 7.55 & - & - \\
44 & NGC 3983        & 179.0986 & 23.8679 & 0.0140 & 10.55 & 5.68 & -11.98 & -3 & $<$ 7.78 & - & - \\
45 & LEDA 46683      & 200.5125 & 35.5139 & 0.0192 & 10.54 & 5.82 & -12.83 & -5 & 8.32$\pm $ 0.08 & 7.6 & 212$\pm $96 \\
46 & LEDA 48265      & 204.7461 & 31.3651 & 0.0153 & 10.34 & 5.83 & -13.05 & -2 & $<$ 7.29 & - & - \\
47 & LEDA 49168      & 207.7509 & 36.9544 & 0.0174 & 10.53 & 5.53 & -12.48 & -5 & 8.32$\pm $ 0.08 & 7.6 & 483$\pm $44 \\
48 & LEDA 2106967    & 208.2216 & 37.6895 & 0.0177 & 10.26 & 5.50 & -12.74 & -5 & $<$ 7.69 & - & - \\
49 & LEDA 50381      & 211.8553 & 39.7475 & 0.0186 & 10.45 & 5.48 & -12.66 & -2 & 7.84$\pm $ 0.11 & 5.0 & 96$\pm $14 \\
50 & LEDA 50498      & 212.2664 & 17.7662 & 0.0161 & 10.29 & 5.74 & -12.81 & -2 & $<$ 7.51 & - & - \\
51 & NGC 5557        & 214.6072 & 36.4936 & 0.0107 & 10.87 & 5.14 & -12.41 & -5 & $<$ 7.98 & - & - \\
52 & NGC 5553        & 214.6239 & 26.2879 & 0.0150 & 10.48 & 5.56 & -12.38 & -3 & $<$ 7.50 & - & - \\
53 & LEDA 51190      & 214.9029 & 17.6444 & 0.0184 & 10.51 & 5.57 & -11.58 & -2 & 8.51$\pm $ 0.08 & 9.3 & 695$\pm $113 \\
54 & LEDA 51317      & 215.4677 & 23.5170 & 0.0171 & 10.29 & 5.60 & -12.71 & -5 & $<$ 7.92 & - & - \\
55 & LEDA 51359      & 215.6353 & 39.5849 & 0.0170 & 10.24 & 5.09 & -12.21 & -5 & $<$ 7.64 & - & - \\
56 & UGC 9397        & 218.8913 & 36.8185 & 0.0142 & 10.27 & 5.20 & -11.53 & -1 & 8.53$\pm $ 0.07 & 11.0 & 309$\pm $204 \\
57 & LEDA 52363      & 219.8029 & 15.8774 & 0.0183 & 10.32 & 5.14 & -12.54 & -5 & $<$ 7.73 & - & - \\
58 & NGC 5928        & 231.5120 & 18.0736 & 0.0149 & 10.88 & 5.67 & -14.00 & -5 & $<$ 8.15 & - & - \\
59 & UGC 9937        & 234.3454 & 20.5499 & 0.0148 & 10.56 & 6.06 & -12.39 & -2 & $<$ 7.51 & - & - \\
60 & NGC 6003        & 237.3568 & 19.0321 & 0.0135 & 10.41 & 5.61 & -12.69 & -5 & $<$ 7.29 & - & - \\
61 & UGC 10388       & 246.7624 & 16.3822 & 0.0154 & 10.46 & 7.08 & -13.06 & 1 & 9.30$\pm $ 0.05 & 56.2 & 463$\pm $5 \\
62 & UGC 10528       & 251.2062 & 22.5214 & 0.0143 & 10.88 & 5.16 & -11.99 & 1 & 9.82$\pm $ 0.05 & 222.1 & 521$\pm $4 \\
63 & LEDA 60388      & 262.3549 & 24.8822 & 0.0144 & 10.46 & 5.69 & -12.75 & -5 & 7.95$\pm $ 0.08 & 6.9 & 347$\pm $128 \\
64 & LEDA 61177      & 269.4643 & 23.8704 & 0.0195 & 10.39 & 5.18 & -11.81 & -3 & 8.50$\pm $ 0.07 & 10.6 & 289$\pm $21 \\
65 & LEDA 61235      & 269.8929 & 24.8870 & 0.0198 & 10.92 & 5.64 & -12.76 & -5 & 8.97$\pm $ 0.05 & 9.4 & 463$\pm $27 \\
66 & UGC 11197       & 274.5950 & 21.2929 & 0.0174 & 10.86 & 5.55 & -11.48 & -5 & $<$ 7.68 & - & - \\
67 & NGC 6628        & 275.5913 & 23.4786 & 0.0148 & 10.96 & 5.09 & -11.55 & 10 & 9.90$\pm $ 0.05 & 65.7 & 502$\pm $9 \\
68 & LEDA 61825      & 275.9697 & 21.0516 & 0.0159 & 10.19 & 5.10 & -11.44 & -1 & $<$ 7.66 & - & - \\
69 & LEDA 62117      & 279.1408 & 19.6210 & 0.0162 & 10.64 & 6.25 & -12.71 & -5 & $<$ 7.78 & - & - \\
70 & LEDA 62146      & 279.3625 & 18.8460 & 0.0179 & 10.83 & 5.81 & -11.92 & -5 & 9.16$\pm $ 0.05 & 22.9 & 328$\pm $63 \\
71 & LEDA 2812114    & 280.2699 & 20.6280 & 0.0159 & 10.20 & 5.24 & -12.17 & -3 & 8.10$\pm $ 0.1 & 5.8 & 444$\pm $206 \\
72 & LEDA 62276      & 280.5213 & 20.1414 & 0.0146 & 10.62 & 5.99 & -11.97 & -5 & $<$ 8.19 & - & - \\
73 & LEDA 5060985    & 336.5996 & 39.4140 & 0.0182 & 10.44 & 5.85 & -12.04 & 10 & 8.96$\pm $ 0.05 & 26.4 & 386$\pm $21 \\
74 & NGC 7461        & 345.4514 & 15.5825 & 0.0141 & 10.47 & 5.66 & -11.57 & -1 & $<$ 7.40 & - & - \\
75 & NGC 7509        & 348.0892 & 14.6094 & 0.0162 & 10.74 & 5.31 & -11.31 & -5 & 8.51$\pm $ 0.08 & 9.6 & 115$\pm $23 \\
76 & LEDA 169936     & 350.0046 & 27.2133 & 0.0187 & 10.50 & 6.27 & -12.63 & -2 & 8.51$\pm $ 0.07 & 13.2 & 502$\pm $106 \\
77 & NGC 7659        & 351.4820 & 14.2098 & 0.0132 & 10.46 & 5.81 & -12.65 & -3 & $<$ 7.18 & - & - \\
78 & LEDA 71557      & 352.2046 & 29.7353 & 0.0186 & 10.68 & 5.62 & -12.52 & -3 & $<$ 7.83 & - & -
\end{longtable}

\subsection{The \addd\ sample}

The full ATLAS$^{\rm 3D}$ sample \citep{Cappellari2011-ATLAS3D} consists of 260 nearby early-type galaxies (ETGs) within 42 Mpc and brighter than $M_{\rm K} = -21.5$ mag, including 58 galaxies in the Virgo Cluster. Deep \hi\ observations for 166 of these galaxies were obtained with the Westerbork Synthesis Radio Telescope \citep[WSRT;][]{Serra2012-ATLAS3D}, resulting in 53 detections and 113 non-detections, with limits on the \hi\ mass fraction reaching $\log(M_{\textsc{Hi}}/M_\ast) \sim -4$ for the latter. For this work, we consider the 65 ATLAS$^{\rm 3D}$ galaxies that meet our selection criteria ($M_\ast > 10^{10} M_\odot$, $\text{NUV}-r > 5$, and $\rm \log sSFR < -11\ yr^{-1}$), plotted as orange stars in~\autoref{fig:SFR_Mstar}. The stellar masses for the ATLAS$^{\rm 3D}$ sample, $M_{\rm JAM}$, were derived using the Jeans Anisotropic Multi-Gaussian Expansion method \citep{Cappellari2013} and exhibit a systematic offset of approximately 0.3 dex from NSA-derived masses. To account for this, we adopt $M_{\rm JAM} - 0.3$ as the stellar mass for this sample. Star formation rates are derived from polycyclic aromatic hydrocarbon (PAH) luminosities \citep{Kokusho2017}. 
This sample comprises 16 Virgo Cluster members and 49 field galaxies. In what follows, we refer to this sample of 65 galaxies as the `ATLAS$^{\rm 3D}$' sample.

\begin{figure*}[ht!]
    \centering
    \includegraphics[width=0.47\linewidth]{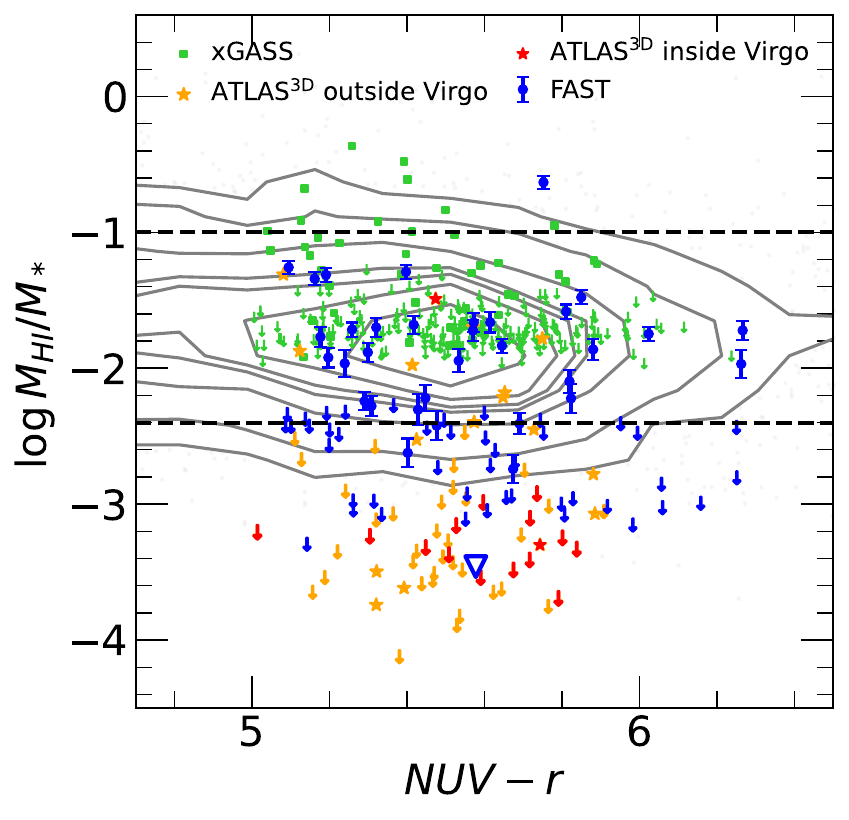}
    \hspace{0.2cm}
    \includegraphics[width=0.47\linewidth]{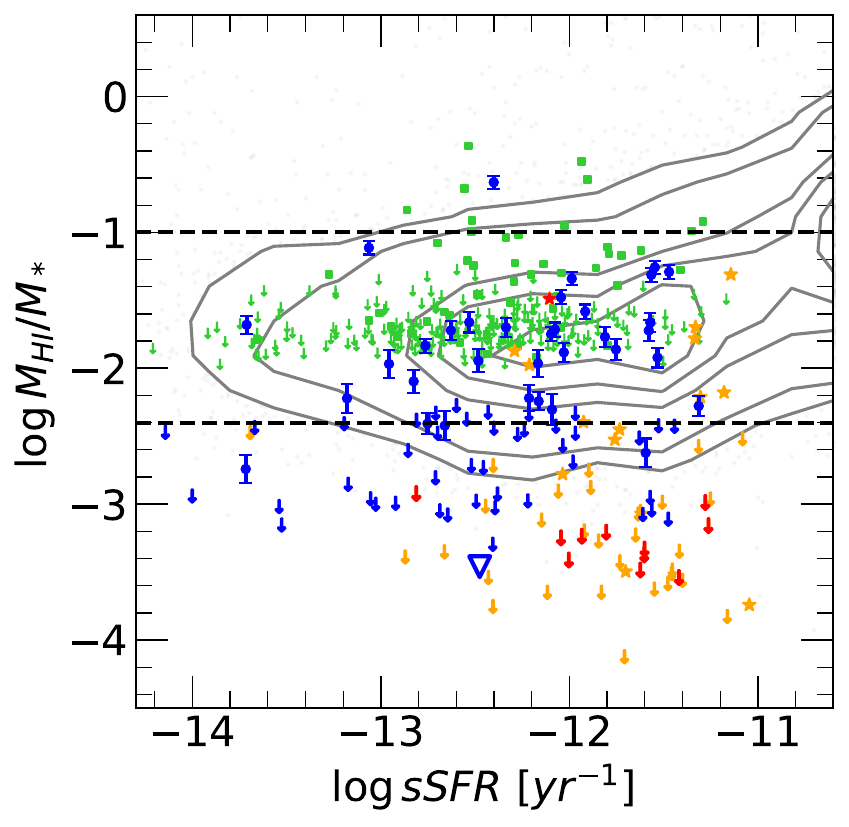}
    \caption{\hi\ fraction as a function of NUV$-r$ and specific star formation rate. The gray contours represent the SDSS volume-limited sample ($M_* > 10^{10} M_{\odot}$).  The colored symbols indicate the \hi\ samples (xGASS: green dots; FAST: blue dots with error bars; \addd\ in the Virgo cluster: red stars; \addd outside the Virgo cluster: orange stars). Upper limits are denoted as downward arrows. The result of the \hi\ spectral stacking analysis for FAST \hi\ non-detections is represented by the open blue downward triangle. The black horizontal dashed lines indicate $\log M_{\hi}/M_*=-1$ and $-2.4$.}
    \label{fig:scaling_relations}
\end{figure*}

\subsection{The xGASS sample} \label{subsec:xGASS}
The extended GALEX Arecibo SDSS Survey \citep[xGASS;][]{xGASS} is a targeted \hi\ survey conducted with the Arecibo telescope. It includes the original GASS sample \citep{GASS} and a low-mass extension of GASS. The GASS sample includes galaxies in the redshift range $0.025 < z < 0.05$ with a flat stellar mass distribution between $10^{10} M_\odot$ and $10^{11.5} M_\odot$. These galaxies were randomly selected from a parent sample of $\sim$12,000 galaxies within the overlapping regions of the SDSS Data Release 6 \citep{Adelman_McCarthy_2008}, the GALEX Medium Imaging Survey \citep{Martin_2005}, and the ALFALFA footprint. Observations continued until either an \hi\ detection was made or a stringent upper limit on the gas fraction of $M_{\textsc{Hi}}/M_\ast < 1.5\%$ was reached. The low-mass extension includes galaxies in the redshift range $0.01 < z < 0.02$ with a flat stellar mass distribution between $10^{9} M_\odot$ and $10^{10.2} M_\odot$. 
The final xGASS representative sample comprises 1179 galaxies representative of the general population. From this sample, we selected galaxies meeting our criteria: $M_\ast > 10^{10} M_\odot$, $\text{NUV}-r > 5$, $\rm \log sSFR < -11\ yr^{-1}$, and a data quality flag of $\texttt{HIconf\_flag} = 0$ (good, uncontaminated detection) or $-99$ (non-detection). Stellar masses and colors are from the NSA, and star formation rates are estimated from SED fitting using the \texttt{CIGALE} code as done above for the FAST and SDSS samples. This selection yielded 253 galaxies with 44 \hi\ detections and 209 non-detections, plotted as green points in~\autoref{fig:SFR_Mstar}. 
For simplicity, we refer to this as the `xGASS' sample hereafter.

\section{Results} \label{sec:results}
\subsection{The \hi\ Content of Massive Quiescent Galaxies}

Figure~\ref{fig:scaling_relations} presents the \hi\ mass fraction as a function of both NUV$-r$ color (left panel) and sSFR (right panel) for massive quiescent galaxies from our FAST, xGASS, and \addd\ samples. \hi\ detections are shown as filled symbols, while non-detections are represented as downward-pointing arrows indicating upper limits. Different colors and symbol styles distinguish between the three samples. The \addd\ galaxies inside and outside the Virgo Cluster are plotted with different colors. For reference, we plotted the predicted distribution of the SDSS volume-limited sample from previous figures (for a fair comparison, the stellar mass threshold of the volume-limited sample has been updated to $M_* > 10^{10} M_{\odot}$) is shown as gray contours in both panels. For clarity, the display is limited to a zoomed-in region (NUV$-r > 4.7$ or $\rm \log sSFR< -10.6$ yr$^{-1}$) where the quiescent galaxy population resides. We categorize our galaxies as follows: \hi-rich, with $\log(M_{\textsc{Hi}}/M_\ast \geq -1$); \hi-middle, with $-2.4 \leq \log(M_{\textsc{Hi}}/M_\ast) < -1$; and \hi-poor, with $\log(M_{\textsc{Hi}}/M_\ast) < -2.4$. These thresholds, indicated by horizontal dashed lines in the figure, approximately correspond to the 90\% contour of the predicted distribution across the parameter space spanned by our samples.

This figure contains several key results from our FAST sample. First, the \hi\ fraction of our FAST sample spans an exceptionally large range of \hi\ mass fractions. Due to the large number of upper limits, we do not know the exact lower limit of the \hi\ fraction distribution. However, even assuming that the non-detections' true \hi\ fractions are equal to their upper limits, the \hi\ fraction of our FAST sample still spans more than 2 orders of magnitude. The \hi\ detections do not exhibit any clear correlation between \hi\ mass fraction and either NUV$-r$ color or sSFR. Given that non-detections have similar NUV$-r$ color and sSFR as detections but significantly lower \hi\ fractions, this lack of correlation likely holds for the entire FAST sample.
 Second, the 34 \hi-detected galaxies (constituting one-third of our sample) generally follow the predicted distribution for quiescent galaxies based on previous surveys. Third, and most strikingly, approximately two-thirds of the sample galaxies remain undetected in \hi, with very low gas fraction upper limits: $\log(M_{\textsc{Hi}}/M_\ast) \lesssim -2.6$ or even $-3.2$. All these non-detections fall in the \hi-poor category, mostly below the outermost contour of the predicted distribution which includes 95\% of the SDSS volume-limited sample. We have attempted to estimate the average \hi\ mass fraction for these 44 non-detected galaxies by stacking their \hi\ spectra using the approach of \citet{Hu2019}. The stacked spectrum reveals no significant \hi\ emission line, but yields a $3\sigma$ upper limit of $\log(M_{\textsc{Hi}}/M_\ast) < -3.46$ (shown as the blue open triangle in~\autoref{fig:scaling_relations}). This corresponds to an \hi\ mass upper limit of $M_{\textsc{Hi}} < 10^{6.96} M_\odot$ given the average stellar mass [$\log(M_\ast/M_\odot)=10.42$] of all the non-detected galaxies in our FAST sample.

Due to its shallower sensitivity, the xGASS sample primarily probes the upper portion of the predicted \hi\ distribution. Despite this limitation, the fact that 83\% of xGASS galaxies are non-detections [$\log(M_{\textsc{Hi}}/M_\ast) \lesssim -1.8$] is consistent with our FAST results, which show a comparably high fraction of galaxies below this limit. The \addd\ sample exhibits a similarly high overall fraction of \hi-poor galaxies. Due to its deeper detection limits and smaller galaxy distances, however, \addd\ shows a significantly greater prevalence of systems below the xGASS detection limit. Furthermore, we find that $86\% \pm 11\%$ (56/65) of \addd\ galaxies fall in the \hi-poor category, compared to $63\% \pm 7\%$ (49/78) for the full FAST sample (errors represent Poisson uncertainties). 

As is well established for the general galaxy population, \hi\ content strongly correlates with optical morphology, largely because morphology itself is tightly linked to star formation activity \citep[e.g.,][]{Fabello2011,Brown2015,Namiki2021,Castignani2022}.
In this work, given the \addd\ sample's exclusive composition of early-type galaxies, the difference between the FAST and \addd\ samples suggests that the morphology-\hi\ content relation still exists even when star formation rate is constrained to be low. When we restrict our FAST sample to the 71 galaxies also classified as early-types, the \hi-poor fraction increases to $68\% \pm 8\%$ (48/71). This reduces the deviation from the \addd\ fraction to $1\sigma$, indicating that morphology plays a role in driving \hi\ poverty, but that another factor may be amplifying the effect in the \addd\ sample. That additional factor is likely environment. Sixteen of the 65 \addd\ galaxies are Virgo Cluster members. For the remaining 49 non-Virgo \addd\ galaxies, the \hi-poor fraction remains high at $84\% \pm 12\%$ (41/49), still significantly exceeding the value for our morphologically-matched FAST sample. We will explore the dependencies on both morphology and environment in detail in the following subsections.



\begin{figure*}[ht!]
    \centering
    \includegraphics[height=0.47\textwidth]{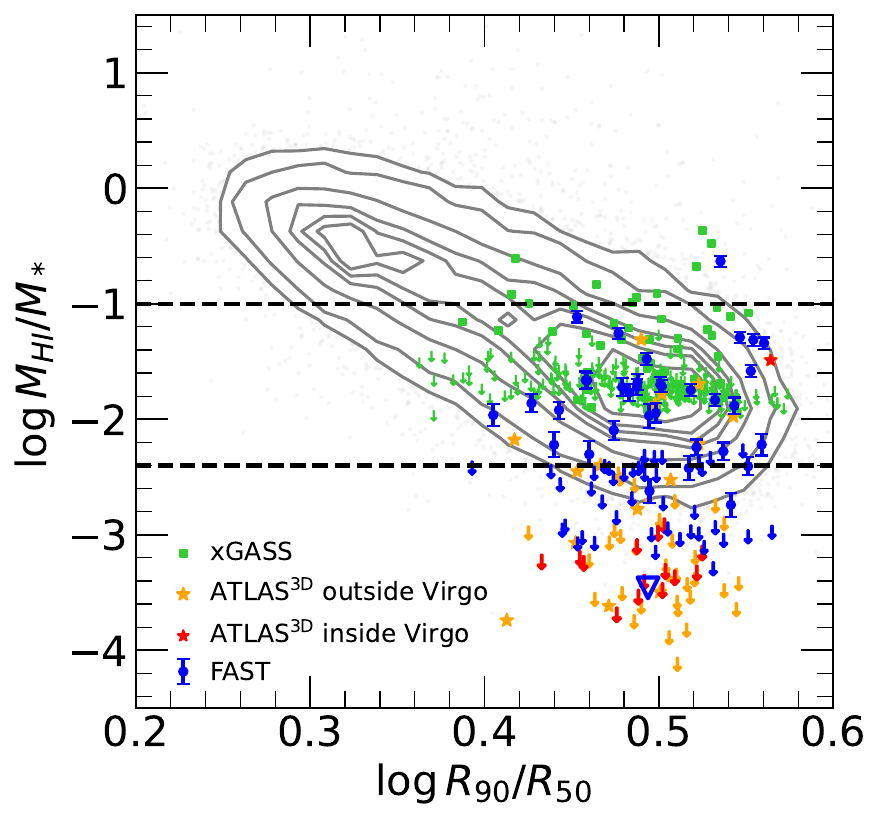}
    \includegraphics[height=0.47\textwidth]{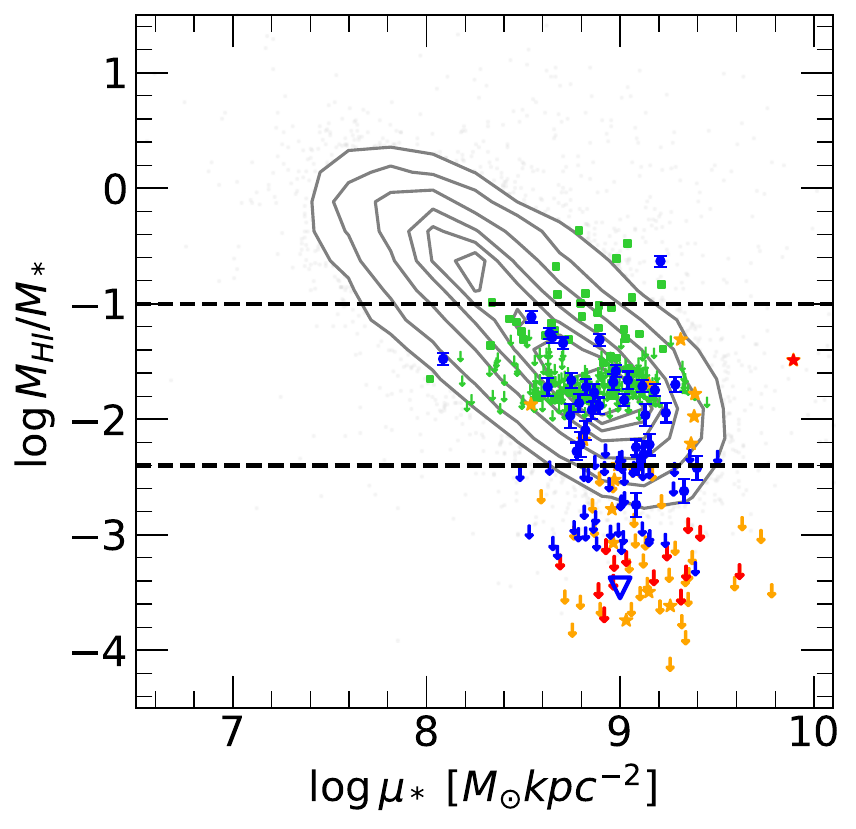}

    \caption{\hi\ fraction as a function of concentration and stellar surface mass density. The gray contours represent the volume-limited sample ($M_* > 10^{10} M_{\odot}$). The FAST sample is denoted as blue points with error bars. The orange(red) symbols indicate \addd\ galaxies in(outside) the Virgo cluster. The xGASS sample is in green. \hi\ non-detections are denoted as downward arrows. The result of the \hi\ spectral stacking analysis for FAST \hi\ non-detections is represented by the open blue downward triangle. The black horizontal dashed lines indicate $\log M_{\hi}/M_*=-1$ and $-2.4$.}
    \label{fig:morphology}
\end{figure*}

\begin{figure*}
    \centering
    \includegraphics[height=0.4\textwidth]{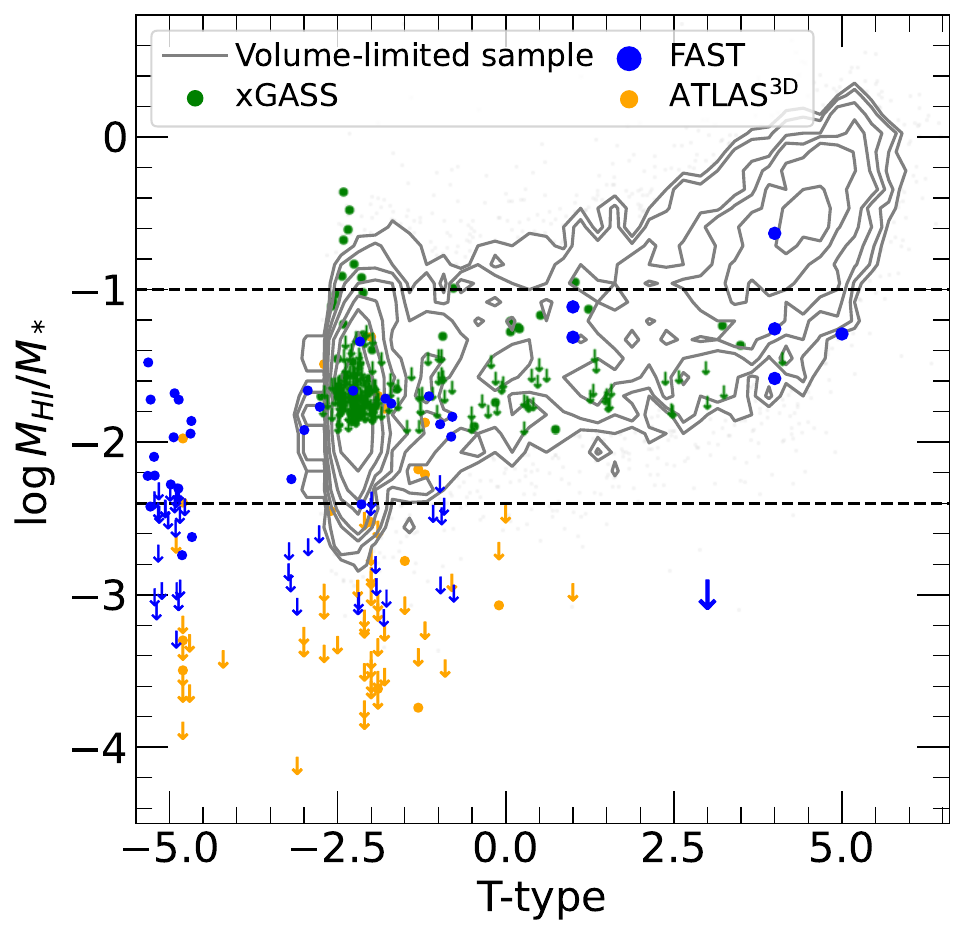}
    \includegraphics[height=0.4\textwidth]{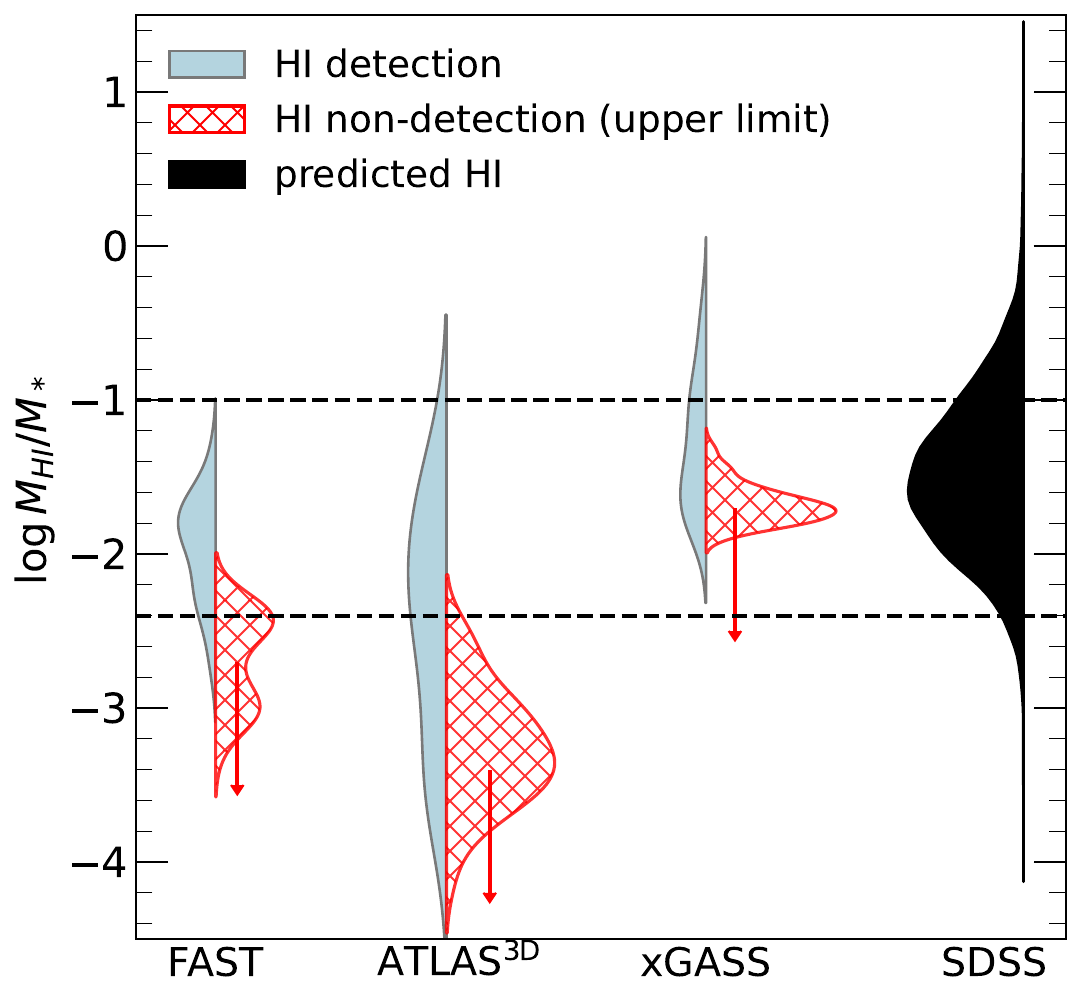}
    \caption{\textbf{Left:} \hi\ fraction as a function of T-type. Colored symbols represent different samples (blue: FAST, orange: \addd, green: xGASS). \hi\ non-detections are shown as arrows. To prevent overlap, the position of each early-type galaxy ($T\leq 0$) in the FAST sample is adjusted with a small horizontal random offset. The gray contours indicate the SDSS volume-limited sample ($M_* > 10^{10} M_{\odot}$). The horizontal dashed lines correspond to $\log M_{\hi}/M_*=-1$ and $-2.4$. \textbf{Right:} \hi\ fraction distribution of early-type galaxies ($T\leq 0$) in the FAST, \addd, xGASS and SDSS volume-limited samples. Real \hi\ observations are colored in lightblue(detection) and red hatch(non-detection). The downward arrows attached to the non-detection distributions serve as a reminder that these are upper limits. Predicted \hi\ fractions are shown in black.}
    \label{fig:ETG_violin}
\end{figure*}

\subsection{Dependence on morphology} \label{result:morphology}


We further investigate the dependence of the \hi\ mass fraction on morphology. We first consider two standard structural parameters: the concentration index $R_{90}/R_{50}$ (where $R_{50}$ and $R_{90}$ are the radii enclosing 50\% and 90\% of the total $r$-band light, respectively) and the stellar mass surface density ($\mu_\ast$). Figure~\ref{fig:morphology} presents $\log(M_{\textsc{Hi}}/M_\ast)$ as a function of these parameters in the left and right panels, respectively. For robust comparison, we include our three \hi\ samples: FAST, xGASS, and \addd, and compare them with the SDSS volume-limited sample for which \hi\ masses are predicted using the estimator from \citealt{XiaoLi}. All $R_{90}/R_{50}$ and $\mu_\ast$ measurements are from the NSA, with different samples shown using the same visual encoding (symbols/colors/line styles) as in the previous figure.

As expected, all samples occupy the region of high concentration indices ($R_{90}/R_{50}$) and high stellar mass surface densities ($\mu_*$), consistent with their massive quiescent nature. Interestingly, no clear trend is observed between \hi\ mass fraction and either parameter. This result shows that massive quiescent galaxies, which tend to be bulge-dominated systems as shown by large concentration indices and stellar mass surface densities, have a significant spread in their \hi\ contents, much more so than disc-dominated galaxies which are overwhelmingly gas-rich \citep{xGASS,Calette2021}.

In addition to structural parameters, we examine the relationship between \hi\ mass fraction and T-type, which provides a quantification of the visual galaxy morphology. The results are presented in the left panel of Figure~\ref{fig:ETG_violin}, using the same visual conventions (symbols/colors/line styles) as previous figures. T-type measurements for the xGASS and SDSS samples are from \citet{Sanchez2018}, who applied a Convolutional Neural Network to SDSS images, while \addd\ T-types are from \citet{Cappellari2011-ATLAS3D}. For our FAST sample, where T-types were unavailable in existing catalogs, we visually inspected DESI images\footnote{\url{https://www.legacysurvey.org}} and classified each galaxy according to the scheme in \citet{Masters2025} (their Table 1). This classification extends to T = -6, contrasting with the S18 scheme which stops at T $\approx$ -2.2. Nevertheless, galaxies with T $\leq 0$ are broadly classified as early-types following common practice. 

By selection, the \addd\ sample contains exclusively early-type galaxies (T $\leq$ 0), with one exception we re-classified as S0 (thus still early-type) following \citet{Masters2025}. While not morphologically selected, the FAST and xGASS samples are dominated by early-types with fractions of $91\%\pm10\%$ and  $85\%\pm 6\%$, respectively. Although limited in number, the seven late-type galaxies in our FAST sample occupy a similar T-type range as late-types in the xGASS sample. These results confirm that our FAST sample exhibits early/late-type fractions and T-type coverage consistent with the larger parent populations. 

Strikingly, we find that nearly all \hi-poor galaxies in both our FAST and \addd\ samples are classified as early-types (T $\leq 0$), strongly suggesting that early-type morphology plays a role in driving \hi\ poverty in nearby galaxies. Conversely, when examining only early-type systems, we find that a substantial fraction in both FAST ($67.6\%\pm 8.0\%$) and \addd\ ($86.2\%\pm 10.7\%$) samples fall within the \hi-poor regime. In contrast, all but one late-type galaxy in our FAST sample reside in the \hi-middle or \hi-rich categories, corresponding to an \hi-poor fraction of $14.3\% \pm 5.4\%$ which is lower than that of early-type galaxies by $5.5\sigma$. This morphological segregation provides further evidence for a morphology-dependent \hi\ content. However, given the small size of the FAST sample, more data are required to confirm this correlation.

The \hi\ content of early-type galaxies across different samples can be more clearly seen in the right panel of Figure~\ref{fig:ETG_violin}, which presents the distribution of $\log(M_{\textsc{Hi}}/M_\ast)$ as vertically-oriented violin plots for all galaxies with T $\leq$ 0. From left to right, the plots correspond to the FAST, \addd, xGASS and SDSS samples. For FAST, \addd\ and xGASS, detections and non-detections are plotted as lightblue and red hatched regions, separately. Note that, for non-detections, the regions represent the distribution of their \hi\ fraction upper limits but not the true \hi\ fraction. In both the FAST and \addd\ samples, non-detections are dominated by \hi-poor galaxies, while detections primarily fall within the \hi-middle category. This distribution is a direct consequence of our operational definition of `\hi-poor'. The double-peaked distribution of non-detections in the FAST sample results from the two different sensitivities employed in our observations. Although the FAST sample has higher detection limits compared to the deeper \addd\ survey, both samples clearly demonstrate that the majority of massive quiescent early-type galaxies are extremely \hi-poor. However, the fact that not all early-type galaxies are \hi-poor indicates that early-type morphology is not a sufficient condition for \hi\ poverty. This suggests that additional physical processes---such as environmental effects---likely contribute to the extreme gas depletion observed in local massive quiescent galaxies.

\begin{figure*}
    \centering
    \includegraphics[width=0.75\linewidth]{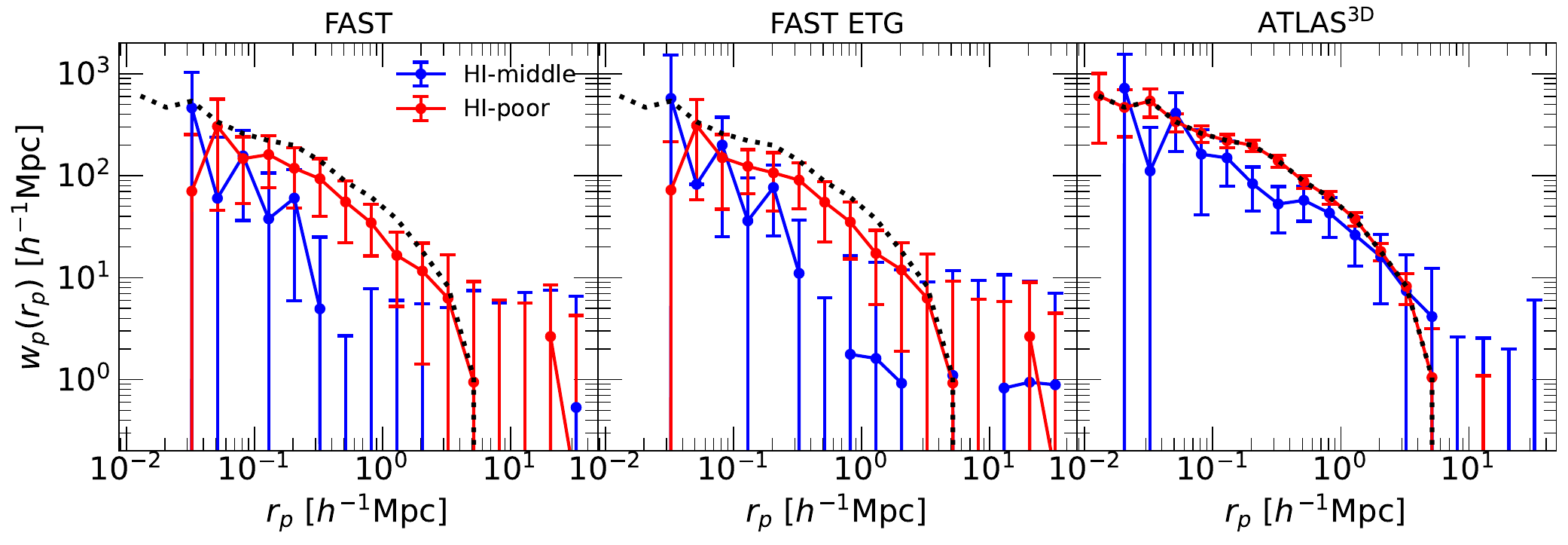}
    \caption{Projected cross correlation function $w_p(r_p)$ of the FAST(left), the FAST early-type galaxy(middle), and the \addd(right) samples) samples. Blue/red solid lines indicate \hi-middle/poor galaxies. The $w_p(r_p)$ of \addd\ \hi-poor galaxies is repeated in every panel as black dotted lines.}
    \label{fig:PCCF}
\end{figure*}

\begin{figure*}
    \centering
    \includegraphics[width=0.75\linewidth]{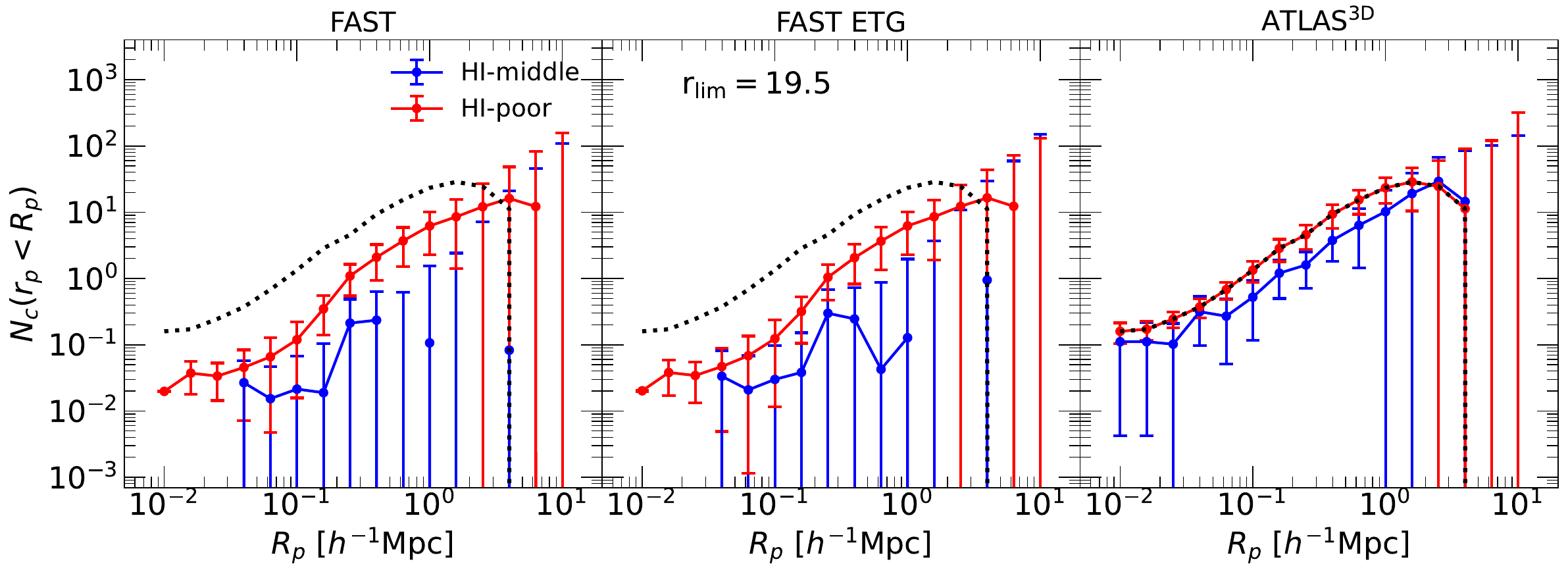}
    \caption{Neighbour counts $N_C(<R_p)$ of the FAST(left), the FAST early-type galaxy(middle), and the \addd(right) samples. The $r$-band magnitude limit is $r_{\rm lim}=19.5$. Blue/red solid lines indicate \hi-middle/poor galaxies. The $N_C(<R_p)$ of \addd\ \hi-poor galaxies is repeated in all corresponding panels as black dotted lines.}
    \label{fig:neighbour_counts}
\end{figure*}

\subsection{Dependence on environment} \label{result:environment}

We now investigate the environmental dependence of \hi\ content in massive quiescent galaxies. Following \citet{Li2006-AGN}, we measure two statistics for a given \hi\ sample: the projected cross-correlation function $w_p(r_p)$ and the background-subtracted neighbour count $N_c(<R_p)$. For $w_p(r_p)$, we cross-correlate each \hi\ sample with a reference sample of $\sim 5.3\times 10^5$ galaxies from the SDSS spectroscopic survey, quantifying clustering from $\sim$10 kpc to $\sim$10 Mpc scales [higher $w_p(r_p)$ means stronger clustering]. The neighbour count $N_c(<R_p)$ measures the average number of neighbouring galaxies within a projected radius $R_p$ around our \hi\ sample galaxies, calculated using the SDSS photometric sample to a depth limit of $r_{\text{lim}} = 19.5$\footnote{We have obtained $N_c$ measurements for two more limits ($r_{\text{lim}} = 18.5, 20.5$), and found our results are similar and robust to the choice of the limiting magnitude.}. This approach probes environments traced by fainter galaxies than possible with the spectroscopic sample ($r_{\lim} = 17.6$). We statistically subtract background/foreground contamination using established methods. Technical details and validation tests for both statistics can be found in \citet{Li2006-AGN, Li2008a, Li2008b, Wang2019}.

In this analysis, we consider three \hi\ samples: the full FAST sample, the early-type subset of the FAST sample, and the \addd\ sample. For each sample, we classify galaxies as \hi-middle or \hi-poor using the established threshold of $\log(M_{\textsc{Hi}}/M_\ast) = -2.4$. We then measure both $w_p(r_p)$ and $N_c(<R_p)$ for each subset. Our measurements of $w_p$ and $N_c$ are presented in Figures~\ref{fig:PCCF} and~\ref{fig:neighbour_counts}, respectively. Across all samples, \hi-poor galaxies consistently exhibit at least marginally stronger clustering amplitudes and higher neighbor counts than \hi-middle galaxies at intermediate scales ($\sim$100 kpc to $\sim$1 Mpc). This finding aligns with earlier clustering studies of \hi-selected samples by \citet{Li2012} and \citet{Guo2017}, though those investigations were limited to relatively \hi-rich galaxies due to shallower survey depths. 

We find that the early-type subset of the FAST sample yields results nearly identical to those of the full sample, albeit with greater noise due to the smaller sample size. This finding is expected, as it reflects the previously established fact that the majority of galaxies in our FAST sample are early-types. When compared to the exclusively early-type \addd\ sample, the early-type galaxies in our FAST sample exhibit weaker clustering amplitudes and lower neighbor counts at fixed scale and magnitude limit ($r_{\text{lim}}$). 
This difference in clustering and neighbour counts, indicative of lower density environments for galaxies in our FAST sample, likely explains their somewhat higher gas fractions, since it is well-established that  galaxies in massive halos are particularly \hi-deficient \citep{Li2012,Serra2012-ATLAS3D,Marasco2016,Brown2017,Guo2017,Stevens2019,Reynolds2022,Cortese2021}.  The specific inclusion of Virgo cluster galaxies in the ATLAS3D explains part of the bias of that sample towards denser environments. Nevertheless, both samples reveal a similar dependence of $w_p$ and $N_c$ on \hi\ mass fraction, consistently underscoring the significant role of environment in regulating the \hi\ content of massive quiescent galaxies, even when restricted to early-type morphologies.

As demonstrated by \citet{Li2006-AGN}, weaker clustering at intermediate scales indicates a higher fraction of central galaxies in a sample. Consequently, the stronger intermediate-scale clustering observed for \hi-poor galaxies in both our FAST and \addd\ samples suggests they preferentially reside as satellites within their host dark matter halos. To quantitatively examine the central/satellite fractions, we cross-match our samples with the SDSS galaxy group catalog from \citet{Yang2007-group-catalog}. For the small fraction of galaxies without counterparts in this catalog, we classified a galaxy as `isolated' (and thus central) if it is more massive than any neighbor within a projected radius of $\Delta r_p < 500$ kpc and a line-of-sight velocity difference of $|\Delta V| < 500$ km/s. Additionally, all \addd\ galaxies within the Virgo Cluster are treated as satellites. The resulting numbers of central and satellite galaxies in the FAST and \addd\ samples, as well as the central fractions and their Poisson errors are listed in~\autoref{tab:morphology_environment}.

\begin{table}[!ht]
  \centering
  \caption{Central/satellite classification of the FAST and \addd\ samples. The numbers represent the galaxy counts in each category. Numbers in the bracket represent early-type galaxies.}
  \label{tab:morphology_environment}
  \begin{tabular}{lcccc}
    \toprule
    \multirow{2}{*}{} & \multicolumn{2}{c}{FAST} & \multicolumn{2}{c}{\addd} \\
    \cmidrule(lr){2-3} \cmidrule(lr){4-5}  
                      & \hi-middle & \hi-poor & \hi-middle & \hi-poor \\
    \midrule
    Central & 24 (19) & 28 (28) & 8 & 29 \\
    Satellite & 4 (4) & 21 (20) & 1 & 27 \\
    $f_{\text{cen}}$ & 86\% (83\%) & 57\% (58\%) & 89\% & 52\% \\
    $\Delta f_{\text{cen}}$ & 16\% (17\%) & 8\% (8\%) & 30\% & 7\% \\
    \bottomrule
  \end{tabular}
\end{table}

We find central fractions of $68\% \pm 8\%$ for the FAST sample and $57\% \pm 7\%$ for the \addd\ sample. When divided into \hi-middle and \hi-poor subsets, the central fractions approach $\sim$90\% for \hi-middle galaxies in both samples, significantly higher than the fractions of $\lesssim 60\%$ found for \hi-poor galaxies. This trend persists when restricting the analysis to early-type galaxies. Although these central fractions are subject to considerable uncertainty due to small-number statistics, the consistent results from both the FAST and \addd\ samples strongly support our conjecture that \hi-poor galaxies have a higher satellite fraction than their \hi-middle counterparts. This finding is physically well-motivated, as it is established that satellite galaxies experience environmental processes such as tidal and ram-pressure stripping, which efficiently reduce their \hi\ content upon infall into groups and clusters \citep[e.g.,][]{Zhang2013,Brown2017,Cortese2021}.

\section{Discussion}\label{sec:discussion}

\subsection{The true \hi\ content of massive quiescent galaxies}

For early-type galaxies, \addd\ has previously demonstrated that most massive quiescent early-type galaxies are very \hi-poor. Our work further extends the investigation to the general population of galaxies in the local Universe. As a representative and deep \hi\ sample of massive quiescent galaxy population at $z=0$, our FAST sample shows that two-thirds of massive quiescent galaxies fall below $\log M_{\hi}/M_* = -2.4$. The remaining one-third basically follows the \hi\ fraction distribution predicted by the \hi\ estimator of \cite{XiaoLi}. These results not only confirm the discovery of \addd\ but also indicate that the massive quiescent population at $z=0$ is dominated by \hi-poor galaxies. In addition, our FAST sample reveals a possible correlation between \hi\ content and optical morphology, as demonstrated by the remarkably different \hi-poor fractions observed in early-type and late-type FAST samples. Given the small sample size (especially for late-type galaxies) of the FAST sample, this correlation is still tentative, and more data are required to confirm this correlation.

Our FAST sample contains only one \hi-rich galaxy, exhibiting an extremely high \hi\ mass fraction of $\log(M_{\textsc{Hi}}/M_\ast) = -0.6$ that places it well above the 95\% contour of the predicted distribution. This result aligns with previous studies \citep[e.g.,][]{Cortese2020, XiaoLi2024}, which have shown that massive quiescent galaxies with unusually high \hi\ content do exist but are exceptionally rare. For instance, \citet{XiaoLi2024} found that the fraction of low-star formation rate but \hi-rich galaxies is $\ll 1\%$ in most cases. 

Our results also reinforce the established consensus that existing \hi\ surveys are either too shallow or morphologically biased, a limitation that is most severe for massive quiescent systems. Given the crucial role of \hi\ gas in galaxy formation and evolution, a complete understanding requires a new generation of deep, large-area \hi\ surveys (e.g. the ongoing one-hundred-square-degree \hi\ deep survey with FAST, Xu et al. in prep). Such efforts must push detection limits from the current $\sim$1\% threshold down to $\sim10^{-3}$ or even $\sim10^{-4}$ in \hi\ mass fraction, enabling comprehensive characterization of the cold gas content across the full parameter space of galaxy populations.

\subsection{Physical mechanisms for the extreme \hi\ poverty}

To understand what causes the extreme low \hi\ gas content for the majority of massive quiescent galaxies, we have investigated the dependence of \hi\ content on both morphology and environment. Our analysis suggests that both early-type morphology and satellite status may contribute to the \hi\ poverty among massive quiescent galaxies, consistent with previous studies. Previous studies have well established that satellite galaxies are subject to a variety of environmental effects within their host dark matter halos. These include ram-pressure stripping \citep[e.g.,][]{Gunn1972-RPS, Abadi1999}, tidal interactions \citep[e.g.,][]{Toomre1972}, harassment \citep[e.g.,][]{Moore1996}, and strangulation \citep[e.g.,][]{Weinman2009}. Collectively, these processes cause satellite galaxies to lose their hot halo gas, cease star formation, and evolve toward earlier morphological types. Consequently, the strong connection between \hi\ poverty, early-type morphology, and satellite status may represent different aspects of the same evolutionary pathway, all fundamentally driven by environmental effects. Visual inspection reveals that both our FAST sample and the \addd\ sample contain numerous S0 galaxies. This provides further evidence for environment-driven morphological transformation, wherein spirals may be transformed into S0s through processes like ram-pressure stripping, which fades their disks and alters their morphology \citep[e.g.,][]{Boselli2006, Barway2009}.

However, our results show that even the combination of early-type morphology and satellite status does not guarantee \hi\ poverty. This indicates that additional physical processes that are currently unidentified must also be responsible for driving the exceptional \hi\ depletion in this population.

For central galaxies in the \hi-poor category, the environmental effects are expected to be negligible. Instead, their extreme \hi\ poverty may be driven by alternative mechanisms. These include: i) inefficient gas cooling due to shock heating in massive dark matter halos \citep[e.g.,][]{Binney1977, Rees1977, Silk1977, Blumenthal1984, Birnboim2003, Keres2005, Dekel2006, Cattaneo2006}; ii) gas heating and ejection due to AGN and stellar feedback  \citep[e.g.,][]{DiMatteo2005, Hopkins2006, Booth2009, Kilborn2009,Dubois2012, Weinberger2017,Dalla_2012,Hopkins2014,Pillepich2018}; iii) major mergers between gas-rich disk galaxies, which can trigger intense, rapid starbursts that consume cold gas efficiently and leave behind a massive early-type remnant \citep[e.g.,][]{Toomre1972}.
For example, by comparing \hi\ mass fractions in IllustrisTNG \citep{Springel_2018} and EAGLE \citep{Crain_2015} simulations with observational data from the xGASS sample, \citet{XiaoLi2025a} demonstrated that stellar feedback plays a more critical role than AGN feedback in regulating the \hi\ content of low-redshift central galaxies, which are potential progenitors of massive quiescent centrals examined in this work.

A comprehensive understanding of the relative importance of these mechanisms in driving extreme \hi\ poverty will require future work that combines detailed analysis of hydrodynamical simulations incorporating these physics with large, multi-wavelength galaxy surveys including deep \hi\ emission data.

\subsection{Implications for \hi\ mass estimators}
\begin{figure}[t!]
    \centering
    \includegraphics[width=0.97\linewidth]{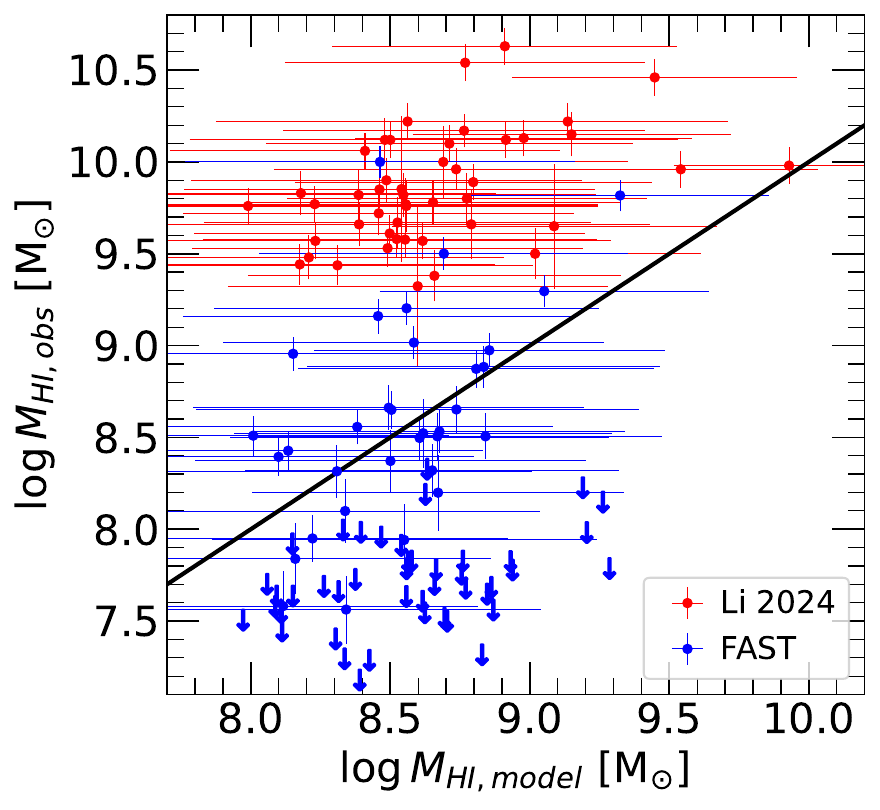}
    \caption{Comparison between the observed \hi\ mass and the predicted \hi\ mass. The quenched \hi-rich galaxies in \cite{XiaoLi2024} are denoted as red points. The FAST sample is in blue (points for \hi\ detections and arrows for \hi\ non-detections). The black solid line indicates the one-to-one line. The error bars indicate the $2\sigma$ uncertainty of the observed and predicted \hi\ mass.}
    \label{fig:model_obs_compare}
\end{figure}

Our FAST observation shows that the majority of the massive quiescent galaxies are more \hi-poor than predicted. One might speculate that the \hi-poor galaxies that fall outside the predicted distribution of the SDSS sample could represent the extended tail of the \hi\ mass distribution. To test this possibility, we perform a direct comparison between observed and predicted \hi\ masses for our FAST sample in~\autoref{fig:model_obs_compare}. As in previous figures, detections are shown as blue points and non-detections as downward-pointing arrows. For each galaxy, we plot the mean predicted \hi\ mass from the estimator of \cite{XiaoLi}, with horizontal error bars indicating the $2\sigma$ uncertainty. For clarity, error bars are shown only for detections, though non-detections exhibit similar uncertainties.

As shown, the \hi\ detections in our sample largely follow the one-to-one relation within the $2\sigma$ confidence interval, consistent with their location within the 95\% contour of the SDSS sample distribution. In contrast, the predicted \hi\ masses for the non-detections are systematically higher than the observed upper limits, with deviations exceeding $2\sigma$. This systematic offset cannot be explained simply by the intrinsic scatter of individual galaxies around the scaling relations underpinning the \hi\ mass estimator. This result confirms that our \hi\ estimator fails to provide reliable predictions for these \hi\ non-detected galaxies.

For comparison, we include the quenched yet \hi-rich galaxies studied by \citet{XiaoLi2024}, plotted as red points with horizontal error bars in~\autoref{fig:model_obs_compare}. These galaxies also exhibit significant deviations from the one-to-one relation, though in the opposite direction, possessing much higher \hi\ masses than predicted. This reinforces the conclusion from \citet{XiaoLi2024} that these \hi-rich quenched galaxies represent a unique, albeit rare, population.

Both \addd\ and FAST samples show that the scatter in $\log M_{\hi}/M_*$ for galaxies with high stellar mass, red color, and high concentration is significantly larger than for star-forming, lower-mass, disc-dominated galaxies. For the latter, an \hi\ estimator with a Gaussian likelihood works very well, capturing the observed distribution as demonstrated by \citet{XiaoLi}. However, applying the same likelihood to massive quiescent galaxies does not work, as evidenced by the fact that the majority of galaxies have gas fractions lower than the predicted distribution. While the \addd\ sample first indicated this, our FAST observation provides a clearer and more quantitative confirmation. 
In the future, a new functional form of scatter is required to improve the performance of the \hi\ estimator for massive quiescent galaxies. Furthermore, explicitly incorporating visual morphology and environmental parameters may also help, as suggested by our analysis.

\section{Summary} \label{sec:summary}

To determine the true \hi\ gas content of massive quiescent galaxies in the local universe, we conducted deep \hi\ observations with FAST for a representative sample of 78 galaxies (7 late-type galaxies and 71 early-type galaxies) selected by stellar mass ($M_* > 10^{10} M_\odot$), color (NUV$-r > 5$), and specific star formation rate ($\rm \log sSFR < -11\ yr^{-1}$). We performed two rounds of FAST observation, firstly with a depth down to $\lgfhi=-2.6$ for the whole sample, and then to $\lgfhi=-3.2$ for 19 galaxies that were non-detected in the first round. In total we obtain 34 \hi\ detections and 44 non-detections. We present the results for our FAST sample and compare them with: (1) the \addd\ sample, deep but restricted to early-type morphologies; and (2) the xGASS sample, representative but with shallower \hi\ sensitivity. We systematically examined the dependence of \hi\ mass fraction on various physical parameters, including: star formation indicators (NUV$-r$ and specific star formation rate), structural parameters (concentration index $R_{90}/R_{50}$ and stellar mass surface density $\mu_*$), morphological classification (T-type), and environmental metrics (projected cross-correlation function $w_p(r_p)$, neighbor count $N_c(<R_p)$, and central/satellite fractions). In addition, we have compared the various \hi\ samples with a volume-limited SDSS sample with \hi\  masses predicted using the estimator from \citet{XiaoLi}, in order to test the predictions for the massive quiescent population.

Our main results can be summarized as follows:
\begin{itemize}
    \item The \hi\ mass fraction in massive quiescent galaxies in the local Universe exhibits an extremely wide dynamic range in gas content (spanning over three orders of magnitude), with the majority of the galaxies being extremely \hi-poor. Approximately two-thirds of our FAST sample exhibit \hi\ mass fractions below $\log M_{\hi}/M_*=-2.4$. The remaining third have higher \hi\ gas fractions consistent with other previously \hi-detected massive quiescent galaxies in the xGASS survey with $-2.4 \lesssim \log(M_{\textsc{Hi}}/M_*) \lesssim -1$. Only a single galaxy in our sample is classified as \hi-rich ($\log(M_{\textsc{Hi}}/M_*) > -1$).
    \item Similar to the early-type-only \addd\ sample, early-type galaxies in FAST show a high fraction of \hi-poor populations. In addition, the \hi-poor galaxies in FAST are predominantly early-types. This indicates that early-type morphology may be an important, but not sufficient, condition for the extremely suppressed \hi\ reservoir in massive quiescent galaxies.
    \item The \hi-poor galaxies in both the FAST and \addd\ samples show enhanced clustering amplitudes and higher neighbor counts at intermediate scales ($100 \ \mathrm{kpc} \lesssim r_p \lesssim 1 \ \mathrm{Mpc}$), indicating a higher fraction of satellite galaxies compared to the \hi-normal samples. The high satellite fraction is confirmed by cross-matching with the SDSS group catalog. While satellite status may play a role, it is neither sufficient nor necessary, as a significant fraction of \hi-poor galaxies are central galaxies.
    \item Our results indicate that both early-type morphology and satellite status serve as potential drivers of the extreme \hi\ poverty, consistent with previous studies. However, they alone cannot fully account for it. This compellingly suggests that additional physical mechanisms---currently unidentified---must play a significant role in driving gas depletion in the majority of massive quiescent galaxies.
    \item Only one-third of the massive quiescent galaxies in our FAST sample follow the \hi\ mass fraction distribution predicted by the \hi\ estimator of \cite{XiaoLi} based on previous surveys. This result implies that a new functional form of scatter rather than a Gaussian likelihood would be required to improve the performance of the \hi\ estimator for massive quiescent galaxies, while explicitly incorporating morphological and environmental parameters may also help.
\end{itemize}

\begin{acknowledgments}
We are grateful to the anonymous referee for the helpful comments. X. Li thanks Niankun Yu, Pei Zuo, and Yinghui Zheng for their helpful discussions on FAST data reduction. This work is supported by the National Key R\&D Program of China (grant NO. 2022YFA1602902), the National Natural Science Foundation of China (grant Nos. 12433003, 11821303, 11973030), and China Manned Space Program with grant no. CMS-CSST-2025-A10.
This work has made use of the following software: Numpy \citep{Numpy}, Scipy \citep{SciPy}, Matplotlib \citep{Matplotlib}, Astropy \citep{Astropy2013,Astropy2018,Astropy2022}, Seaborn \citep{seaborn}, corner \citep{corner}.

This work has used the data from the Five-hundred-meter Aperture Spherical radio Telescope (FAST). FAST is a Chinese national mega-science facility, operated by the National Astronomical Observatories of Chinese Academy of Sciences (NAOC).

Funding for the SDSS and SDSS-II has been provided by the Alfred P. Sloan Foundation, the Participating Institutions, the National Science Foundation, the U.S. Department of Energy, the National Aeronautics and Space Administration, the Japanese Monbukagakusho, the Max Planck Society, and the Higher Education Funding Council for England. The SDSS Web Site is http://www.sdss.org/.

The SDSS is managed by the Astrophysical Research Consortium for the Participating Institutions. The Participating Institutions are the American Museum of Natural History, Astrophysical Institute Potsdam, University of Basel, University of Cambridge, Case Western Reserve University, University of Chicago, Drexel University, Fermilab, the Institute for Advanced Study, the Japan Participation Group, Johns Hopkins University, the Joint Institute for Nuclear Astrophysics, the Kavli Institute for Particle Astrophysics and Cosmology, the Korean Scientist Group, the Chinese Academy of Sciences (LAMOST), Los Alamos National Laboratory, the Max-Planck-Institute for Astronomy (MPIA), the Max-Planck-Institute for Astrophysics (MPA), New Mexico State University, Ohio State University, University of Pittsburgh, University of Portsmouth, Princeton University, the United States Naval Observatory, and the University of Washington.

\end{acknowledgments}

\bibliography{refs}{}
\bibliographystyle{aasjournalv7}

\appendix
\section{FAST \hi\ Observation} \label{app:data_reduction}
Our \hi\ observation consists of two parts. First, we observed the \hi\ gas of the 78 selected galaxies in September 2023 and June 2024 (PI: Xiao Li, project code: PT$2023\_0025$). The targets were observed with the ON-OFF mode using the central beam (M01). For 74 target galaxies, the on-source integration time was designed to achieve an \hi\ fraction upper limit of $\log \mhi/\mstar = -2.6$ ($3\sigma$, assuming a velocity width of $W=300\rm\ km\ s^{-1}$). The remaining 4 galaxies reached a deeper \hi\ fraction upper limit of $\log \mhi/\mstar = -3.2$ ($3\sigma, W=300\rm\ km\ s^{-1}$). 
The OFF positions were selected with an angular separation of 5--10 arcmin from their respective targets, ensuring no galaxies with $\Delta \theta < 3'$ and $|\Delta v|<500\rm\ km\ s^{-1}$, where $\Delta \theta$ is the projected angular distance to the OFF position and $|\Delta v|$ is the radial velocity offset relative to the target galaxy.
In each ON-OFF cycle, we injected a high-amplitude ($10\ \rm K$) noise diode for 2s at the beginning of the on-source and the off-source scan. The scan length varies galaxy by galaxy, with a maximum of 300s. This observation yielded 31 \hi\ detections ($3\sigma$ level) and 47 \hi\ non-detections. 

Subsequently, we conducted follow-up observations of 19 \hi\ non-detections from September to October 2024 (PI: Xiao Li; project code: $\rm PT2024\_0250$), aiming for a deeper \hi\ fraction upper limit of $\log \mhi/\mstar = -3.2$ ($3\sigma$, $W=300\rm\ km\ s^{-1}$). These observations employed the ON-OFF mode, covering each target with the central beam (M01) and one outer beam (M10, M12, or M14) in turn during the ON-OFF cycle. The length of each on-source and off-source scan is no longer than 300s. For calibration, a high-amplitude ($10\ \rm K$) noise diode signal was injected for 1s every 64s. We detected \hi\ emission ($3\sigma$ level) in 3 of the 19 target galaxies.

\begin{figure*}[ht!]
    \centering
    \includegraphics[width=0.98\linewidth]{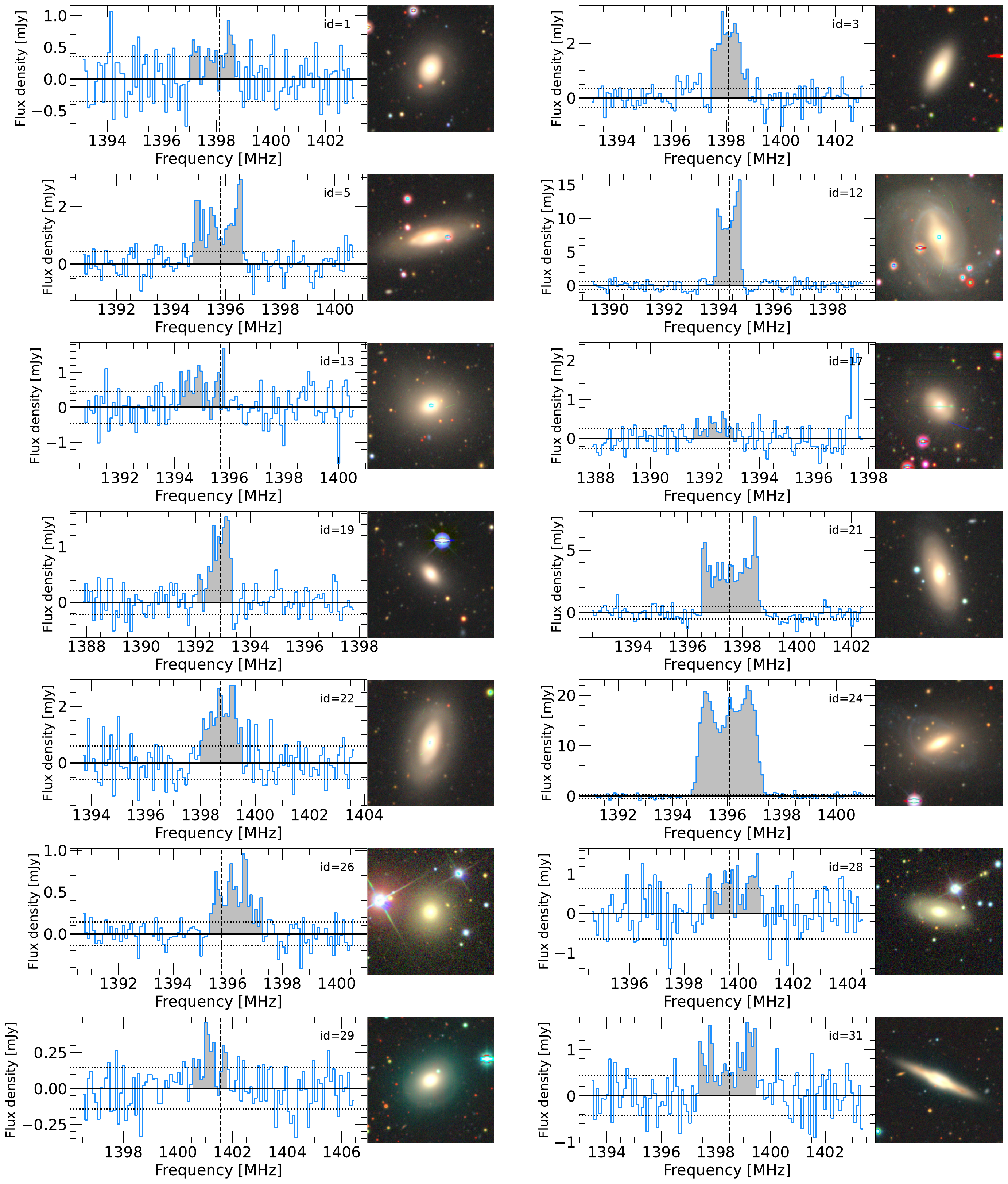}
    \caption{\hi\ spectrum (left sub-panel) and optical image (right sub-panel) of \hi-detected galaxies in the FAST sample. The frequencies are heliocentric. The blue solid line represents the baseline-subtracted \hi\ spectrum. The vertical black dashed line marks the expected frequency of the HI 21cm line derived from optical redshift. The horizontal black dotted line indicates the rms of the spectrum. The gray shaded region shows the spectrum used for integrated flux measurement. The optical image is from the DESI Legacy Imaging Surveys.}
    \label{fig:HIspec_detection}
\end{figure*}

\addtocounter{figure}{-1}
\begin{figure*}[ht!]
    \centering
    \includegraphics[width=0.98\textwidth]{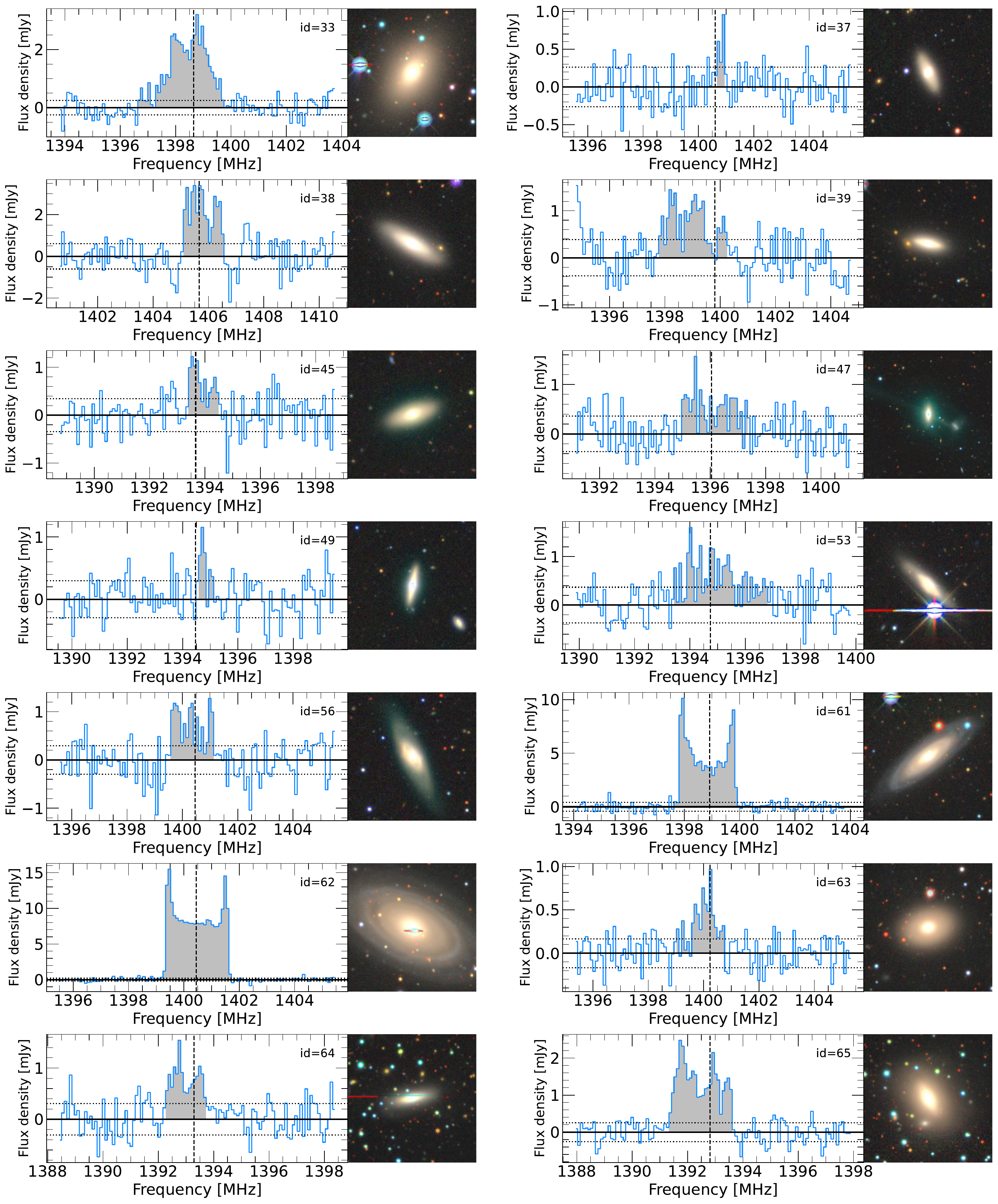}
    \caption{Continued.}
\end{figure*}
\vspace{1cm}

\addtocounter{figure}{-1}
\begin{figure*}[ht!]
    \centering
    \includegraphics[width=0.98\textwidth]{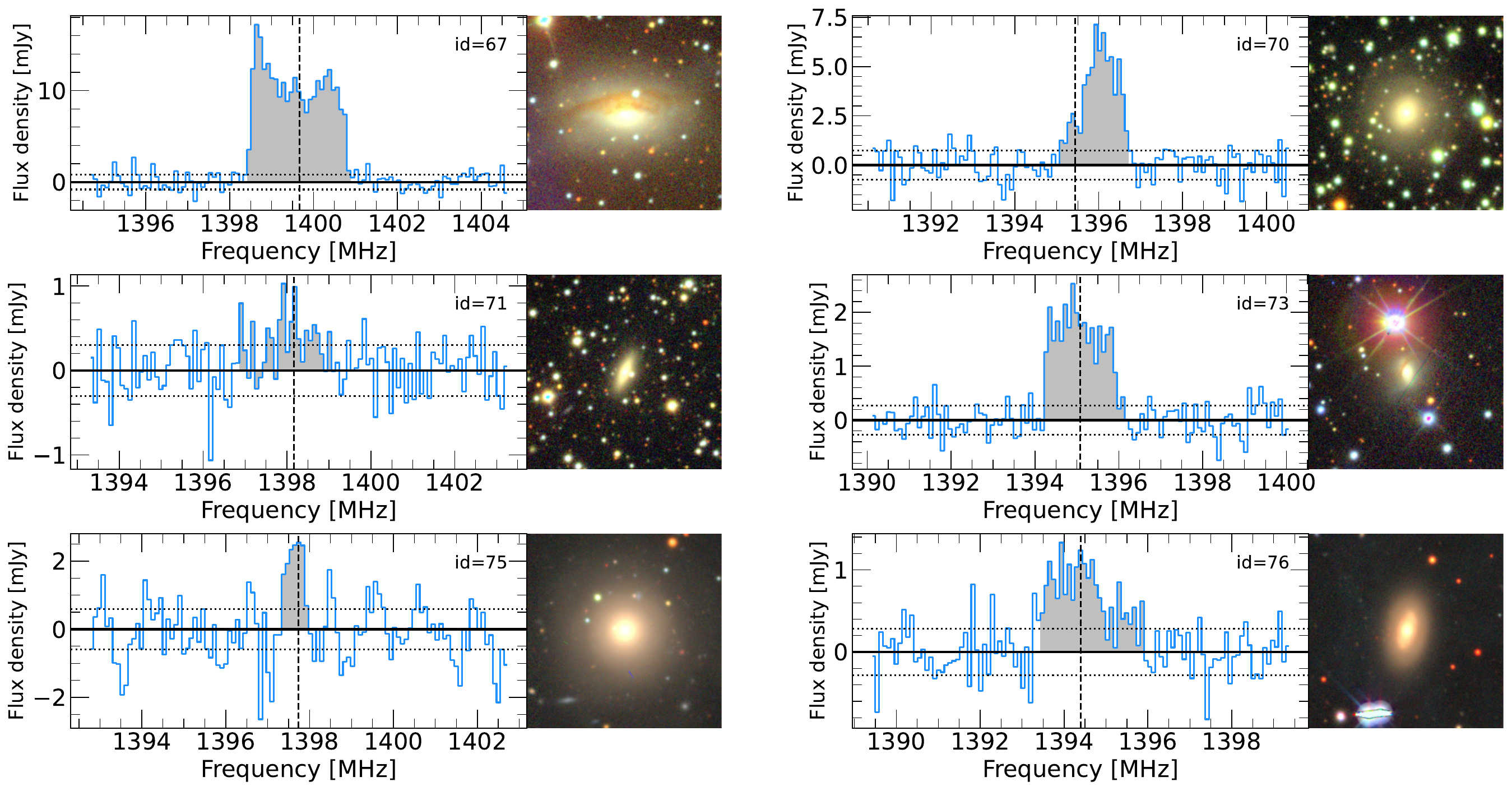}
    \caption{Continued.}
\end{figure*}

The data reduction was performed using our own \texttt{Python} code following the standard ON-OFF mode data reduction procedures. Details of data reduction and calibration can be found in \cite{ChengCheng2020}. Basically, flux calibration was performed following the two equations below:

\begin{equation}
        T_{\rm ON}^{\rm cal\ off} = \frac{T_{\rm noise}}{P_{\rm ON}^{\rm cal\ on} - P_{\rm ON}^{\rm cal\ off}}P_{\rm ON}^{\rm cal\ off}
    \end{equation}
    \begin{equation}
        T_{\rm OFF}^{\rm cal\ off} = \frac{T_{\rm noise}}{P_{\rm OFF}^{\rm cal\ on} - P_{\rm OFF}^{\rm cal\ off}}P_{\rm OFF}^{\rm cal\ off}
    \end{equation}
where $P_{\rm ON}^{\rm cal\ on}$ and $P_{\rm OFF}^{\rm cal\ on}$ refer to the power of on-source and off-source scans with noise diode on. Accordingly, $P_{\rm ON}^{\rm cal\ off}$ and $P_{\rm OFF}^{\rm cal\ off}$ refer to the power of on-source and off-source scans with noise diode off. The power of all scans in the above equations has been time-averaged to one sampling unit. 
$T_{\rm noise}$ is the known temperature of the noise diode. The $T_{\rm ON}^{\rm cal\ off}$ and $T_{\rm OFF}^{\rm cal\ off}$ are the calibrated antenna temperatures of the on-source and off-source spectrum (without noise diode). The calibrated temperatures  are then converted into flux density ($S$) in units of Jansky (Jy) by dividing them by the gain parameters provided in the Table 5 of \cite{Jiang2020-FAST}. Radio frequency interference (RFI) is flagged and masked using sigma clip method.

The spectra of all on-source (or off-source) scans without noise diode are stacked into one spectrum $S_{\rm ON}$ (or $S_{\rm OFF}$) weighted by the inverse square of their rms. The ON-OFF spectrum is $S_{\rm ON-OFF}=S_{\rm ON} - S_{\rm OFF}$. The baseline ripple was then modeled and subtracted by fitting a linear function plus a sinusoidal function. For each source, we visually checked the fitting result to ensure reasonable fitting results.

The original spectrum has a frequency resolution of 7.6 kHz, corresponding to a velocity channel of $dV'\sim 1.7\ {\rm km\ s^{-1}}$ at 1.4 GHz. We re-sample the original spectrum into $dV\sim 20\ {\rm km\ s^{-1}}$, and measure the integrated \hi\ flux ($S_{21}$) by summing up all the velocity channels within the full width at zero intensity (FWZI), which includes the channels at both line wings with the channel flux higher than the rms.

\begin{equation}
    S_{21} = \int_{FWZI}S(V)dV
\end{equation}

Here $S(V)$ is the flux density. The uncertainty of the integrated flux is estimated as $\sigma_{S_{21}} = \sqrt{N_{\rm channel}} \ dV \times \mathrm{rms}$, where $N_{\rm channel}$ is the number of channels used to compute $S_{21}$. The signal-to-noise ratio is then given by $\mathrm{SNR} = S_{21} / \sigma_{S_{21}}$. We adopt $\mathrm{SNR} > 3$ as the criterion for \hi\ detection.
The \hi\ mass is derived using the following formula \citep{Meyer2017}:
    \begin{equation}
        \frac{M_{\hi}}{M_{\odot}}=2.35\times10^5 \left( \frac{D}{\rm Mpc} \right)^2 \left( \frac{S_{21}}{\rm Jy\ km/s} \right)
    \end{equation}
where $D$ is the comoving distance of the source. The error of \hi\ mass is calculated as 

\begin{equation}
    \sigma_{\log \rm M_{\hi}} = \sqrt{(\sigma_{S_{21}}/S_{21})^2\times (1+\sigma_{\rm c})^2 + 0.1^2 + \sigma_{\rm b}^2}\ /\ln{10}
\end{equation}
where the term $0.1$ accounts for the uncertainty of flux calibration. The uncertainty due to baseline subtraction, denoted as $\sigma_{\rm b}$, is tested by fitting the baseline using different spectral windows and fitting functions. According to the test results,
we set $\sigma_{\rm b}=0.05$ for \hi\ detections with $\rm SNR \geq 20$ and to $\sigma_{\rm b}=0.1$ for those with lower SNR. Finally, $\sigma_c$ represents the uncertainty contributed by the radio continuum and is estimated as $\sigma_{\rm c} = T_{\rm c}/T_{\rm sys}$, where $T_{\rm sys}=20$ K is the system temperature of FAST, and $T_{\rm c}=S_{\rm c}\times G$ is the antenna temperature of the radio continuum source in the FAST beam. $G=16\rm\, K/Jy$ is the gain of FAST. $S_{\rm c}$ is either the flux density of the radio continuum source within the FAST beam or --- if no radio continuum source falls within the beam --- the sensitivity of the VLASS survey, which is 0.12 mJy at 2 GHz. We converted this sensitivity to 1.4 GHz assuming a spectral index $\alpha = -0.7$ \citep{VLASS_QL_catalog}.

For \hi\ non-detections, the $3\sigma$ upper limit of \hi\ mass is estimated following \cite{HIMaNGA2}:
    \begin{equation}
    \begin{aligned}
        \frac{M_{\hi,\rm lim}}{M_{\odot}}&=3\times2.35\times10^5 \left( \frac{D}{\rm Mpc} \right)^2 \left( \frac{rms}{\rm Jy} \right) \times \\
        &\sqrt{\left(\frac{W}{\rm km\ s^{-1}}\right)\left(\frac{d V}{\rm km\ s^{-1}}\right)}
    \end{aligned}
    \end{equation}
    where $W$ is the line width, which we assume $W=300\ {\rm km\ s^{-1}}$.
\autoref{fig:HIspec_detection}  show the baseline-subtracted \hi\ spectra and optical images of \hi\ detections. The shaded region in each spectrum marks the frequency range used for integrated flux measurements.

\begin{figure}[ht]
    \centering
    \includegraphics[width=0.45\linewidth]{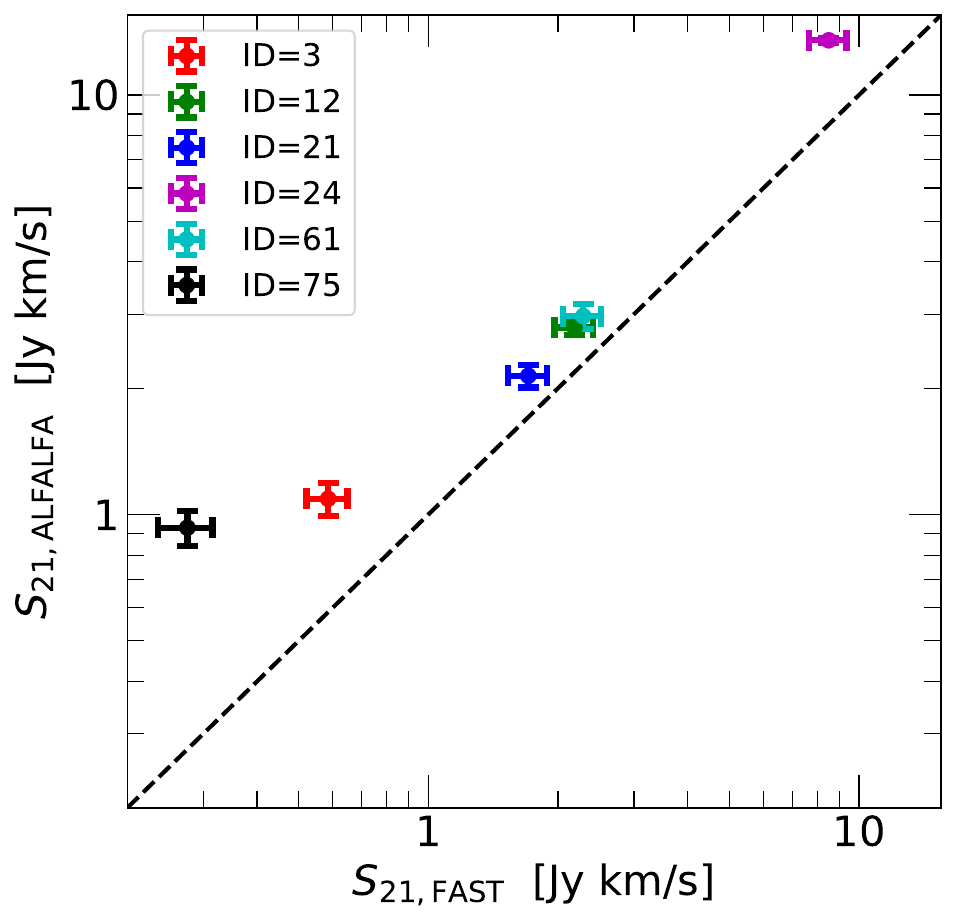}
    \includegraphics[width=0.48\linewidth]{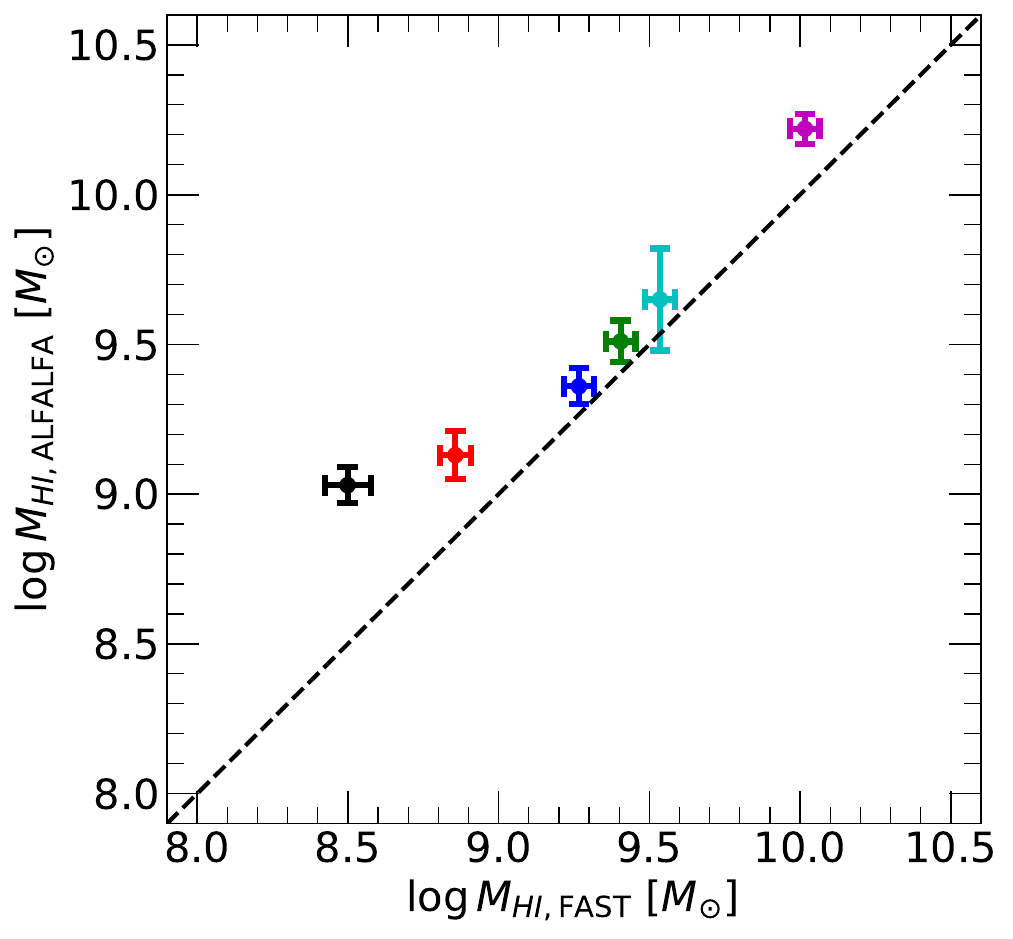}
    \caption{Comparison of \hi\ integrated flux (left) and \hi\ mass (right) measured by FAST and ALFALFA. The legend shows the id of each galaxy. The black dashed line indicates the one-to-one line.}
    \label{fig:compare_HIflux}
\end{figure}

\begin{figure*}[ht]
    \centering
    \includegraphics[width=\linewidth]{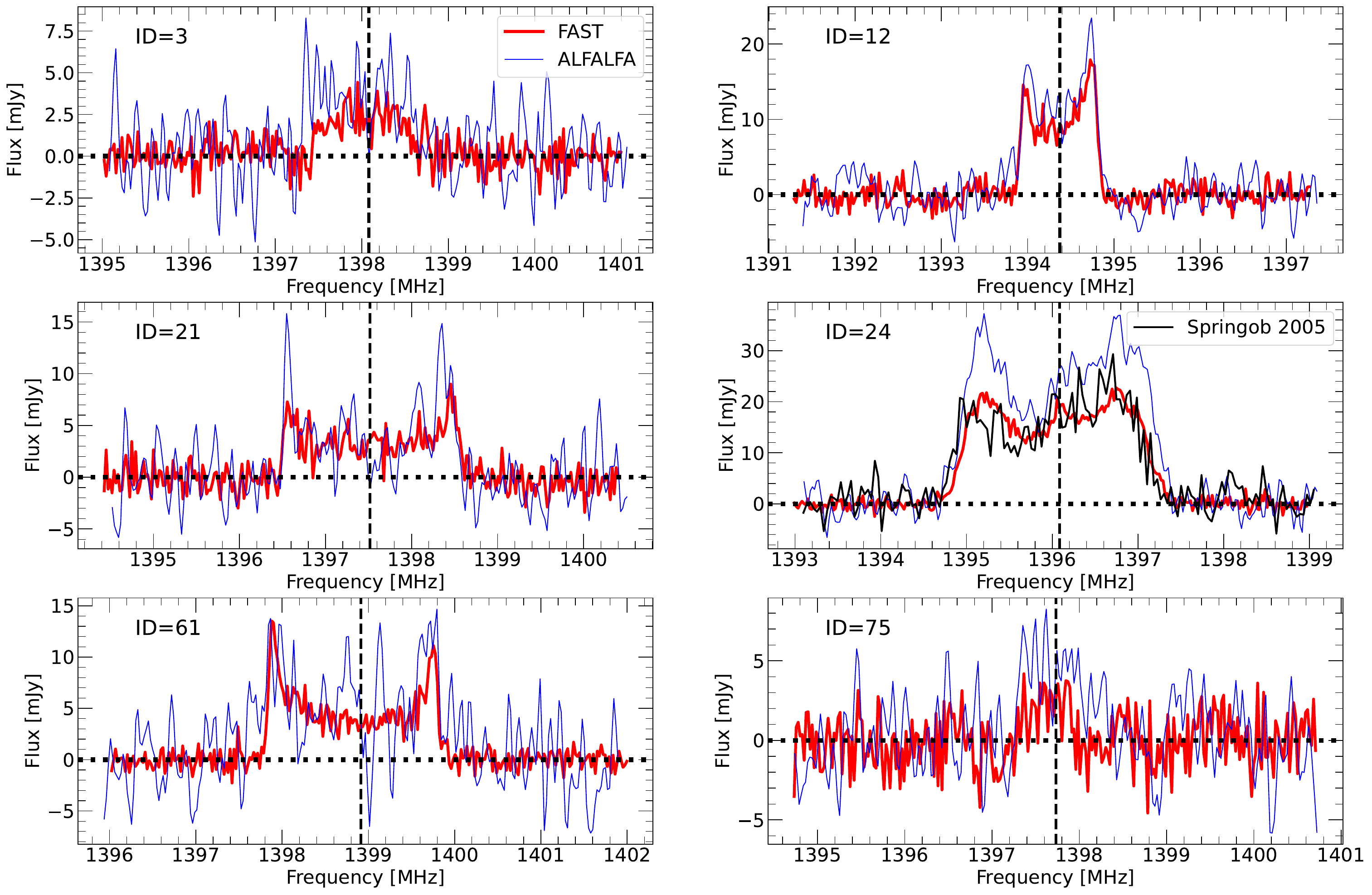}
    \caption{Comparison of \hi\ spectrum measured by FAST(red) and ALFALFA(blue). The frequencies are heliocentric. The vertical dashed line marks the expected frequency of the HI 21cm line derived from optical redshift. The black solid line in the middle-right panel represents the \hi\ spectrum measured in a single Arecibo beam \citep{Springob2005-HIarchive}.}
    \label{fig:compare_HIspectrum}
\end{figure*}

For the 6 galaxies that have ALFALFA spectrum, we compare the \hi\ integrated flux, mass, and spectrum between our FAST observations and ALFALFA data in \autoref{fig:compare_HIflux} and \autoref{fig:compare_HIspectrum}. 
The FAST \hi\ spectrum of source 24 is significantly lower than that of ALFALFA. For this source, we found that the angular diameter of the \hi\ disk estimated from the \hi\ size-mass relation \citep{WangJing2016-HIsizemass-relation} using ALFALFA \hi\ mass is larger than the beam size of FAST, indicating that there is a non-negligible fraction of \hi\ gas missed by FAST, leading to the lower \hi\ flux derived from FAST \hi\ spectrum. This explanation is supported by the \hi\ spectrum obtained by Arecibo \citep[black solid line,][]{Springob2005-HIarchive} with a similar beam size as FAST showing comparable amplitude to our FAST spectrum.
Source 75 is flagged with \texttt{Code=2} in the ALFALFA catalog, indicating poor spectrum quality and large measurement uncertainties. Although source 3 has \texttt{Code=1}, its SNR is similarly low to source 75, suggesting comparable data quality concerns.
For source 12, 21, and 61, we find generally comparable spectral profiles between the two datasets, although FAST fluxes are systematically $20\%-30\%$ lower. The systematically lower FAST \hi\ fluxes for these 5 sources (3, 12, 21, 61, 75) remain puzzling. 
Although their estimated \hi\ diameters from the size-mass relation fall within the FAST beam, we cannot rule out potential flux loss from extended \hi\ components beyond the \hi\ radius \citep{WangJing2014-HIradialprofile}, particularly if the galaxies exhibit irregular \hi\ morphologies rather than regular disks.

\end{document}